\documentclass{ieeeaccess}

 \usepackage{booktabs}
\usepackage{subfigure}

\usepackage{longtable}
\usepackage{enumerate}
\usepackage{color}
\usepackage{url}

\usepackage{algorithm}

\usepackage{cite}
\usepackage{amsmath,amssymb,amsfonts}
\usepackage{algorithmic}
\usepackage{graphicx}
\usepackage{textcomp}
\def\BibTeX{{\rm B\kern-.05em{\sc i\kern-.025em b}\kern-.08em
    T\kern-.1667em\lower.7ex\hbox{E}\kern-.125emX}}
\begin{document}

\bibliographystyle{IEEEtran}
\history{}
\doi{}

\title{Quantum image segmentation based on grayscale morphology}
\author{\uppercase{Wenjie Liu }\authorrefmark{1,2}, 
\uppercase{Lu Wang\authorrefmark{3}, and Mengmeng Cui}.\authorrefmark{1}}
\address[1]{School of Computer and Software, Nanjing University of Information Science  and Technology, Nanjing 210044, China  }
\address[2]{Engineering Research Center of Digital Forensics, Ministry of Education,Nanjing University of Information Science  and Technology,
 Nanjing 210044, China}
\address[3]{School of Automation, Nanjing University of Information Science  and Technology, Nanjing 210044, China}
\tfootnote{This work is supported by the National Natural Science Foundation of China (62071240), the Innovation Program for Quantum Science and Technology (2021ZD0302902), and the Priority Academic Program Development of Jiangsu Higher Education Institutions (PAPD).}

\markboth
{Wenjie Liu \headeretal: Quantum image segmentation based on grayscale morphology}
{Wenjie Liu \headeretal: Quantum image segmentation based on grayscale morphology}

\corresp{Corresponding author: Wenjie Liu (email: wenjiel@163.com).}

\begin{abstract}
The classical image segmentation algorithm based on 
grayscale morphology can effectively segment  images with uneven illumination, but with the increase of the  image data, the real-time problem will emerge. In order to solve this problem, a quantum image segmentation algorithm  is proposed in this paper, which can use quantum mechanism to simultaneously perform morphological operations on all pixels in a grayscale image,  and then quickly segment the image into a binary image.  In addition, several quantum circuit units, including dilation, erosion, bottom hat transformation, top hat transformation, etc., are designed in detail, and then they are combined together  to construct
the complete quantum  circuits  for segmenting the NEQR images. For a $2^n \times 2^n$ image with $q$ grayscale levels, the complexity of our algorithm can be reduced to O$(n^2+q)$, which is an exponential speedup than the classic counterparts. Finally, the experiment is conducted on IBM Q to show the feasibility of our algorithm in the noisy intermediate-scale quantum (NISQ) era.
\end{abstract}


\titlepgskip=-15pt

\maketitle

\section{Introduction}
\label{sec:introduction}
\PARstart{T}{he} image processing technology is the basis of computer vision, and it is widely used in many fields. With the improvement of the performance of image acquisition devices, the quality of images is getting better and better, and the amount of data is also increasing, which also requires more computing power while the image is being processed. However, due to the limitation of some factors, the development of classical computers is close to the limit, so it will consume a lot of time when processing large-scale data,  which will lead to the emergence of real-time problem. In recent years, quantum image processing, as an interdisciplinary subject of image processing and quantum computing, has received extensive attention from researchers \cite{Yan2017,Cai2018,Chakraborty2022}. Due to the unique superposition and entanglement properties of quantum computing, quantum image processing can achieve at most an exponential speedup than its classical counterpart, which is  very meaningful. 

Quantum image representation model is the primary task of quantum image processing \cite{Venegas-Andraca2003,Venegas-Andraca2010}. At present,  scholars have done a lot of research in this area and proposed many quantum image representation models. The most commonly used methods can be mainly divided into two types, and one is to encode the image color information into the probability amplitude, for example, the flexible representation of quantum image (FRQI) \cite{Le2011}, the multi-channel RGB images representation of quantum images (MCQI) \cite{Sun2013}, a normal arbitrary superposition state of quantum image (NASS) \cite{Li2014} and quantum probability image encoding representation (QPIE) \cite{Yao2017}. This encoding method can efficiently encode images using fewer qubits, but it also makes it more difficult to retrieve images.  To solve this problem, a novel enhanced quantum representation (NEQR) \cite{Zhang2013} is proposed, which is an encoding method that uses 
some sequences of qubits to encode images, which encodes the  position information and the grayscale value information into three entangled qubit sequences, such that images can be retrieved quickly with few measurements. Therefore, the NEQR model is widely used due to its simplicity of operation. By improving the position information and color information, a series of encoding methods have been proposed, such as an improved NEQR (INEQR) \cite{Jiang2015}, a generalized model of NEQR (GNEQR) \cite{Zhang2015L}, and a novel quantum representation of color digital images (NCQI) \cite{Li2018}.  With the development of quantum image representation models, corresponding quantum image processing algorithms have also developed rapidly, such as geometrical transformation of quantum image \cite{Le2010,Le2011S,Fan2016}, quantum image steganography based on least significant bit (LSB) \cite{Jiang2016}, feature extraction of quantum image \cite{Zhang2015L},  quantum image watermarking \cite{Mogos2009,Iliyasu2012,Jiang2013,Miyake2016,Jiang2015}, 
quantum image filtering \cite{Li2017}, quantum image scaling \cite{Sang2016}, 
quantum image matching \cite{Jiang2016QIM}, 
quantum image edge detection \cite{Zhang2015,Fan2019,Zhou2019,Chetia2021,Liu2022,Chakraborty2022Q}, 
quantum image segmentation \cite{Li2013,Caraiman2014,Caraiman2015,Chakraborty2018,Xia2019,Yuan2020}, etc.

Image segmentation is the basis of image processing. Although classical image segmentation algorithms are relatively mature, quantum image segmentation algorithms are still in their infancy. In 2014, Caraiman et al \cite{Caraiman2014}. proposed a histogram-based quantum image segmentation algorithm to segment the image according to the distribution of pixel grayscale values in the histogram, which can achieve exponential acceleration than the classical algorithm, but there is no specific oracle implementation circuit. One year later, they  \cite{Caraiman2015} proposed a quantum image segmentation algorithm based on a single threshold and gave an oracle circuit. However, due to the high quantum cost, the algorithm could not be simulated in the noise mesoscale quantum (NISQ) era. In 2018, Chakraborty et al. \cite{Chakraborty2018} designed a multiple quantum circuit for threshold based color image segmentation, which is a meaningful research on color image segmentation. In 2019, Xia et al. \cite{Xia2019} designed a quantum comparator and binarized the quantum image using a single threshold. They designed a detailed segmentation circuit, but its complexity is still too high. In 2020, Yuan et al. \cite{Yuan2020} proposed a double threshold quantum image segmentation algorithm based on the quantum comparator they designed. Because of its low quantum cost, it is feasible in this NISQ era.  However, the above research only uses the threshold value to segment the image, and does not consider the influence of illumination, which is invalid for complex image (image with uneven illumination) segmentation.
 In the real world, the influence of uneven illumination on images is everywhere, which brings great challenges to image segmentation. Classical image segmentation algorithms have done a lot of research to solve this problem, but its complexity is very high, and the real-time problem is gradually emerging. The existing quantum image segmentation methods do not consider the impact of illumination on the image, so they cannot segment the image containing uneven illumination, which is inappropriate in practical use. Therefore, a quantum version algorithm is needed to solve this problem.  Morphology is a new image processing theory, and Yuan et al. \cite{Yuan2015Q} first conducted research on quantum morphology image processing. They focused on the dilation and erosion of quantum binary images and grayscale images, but the complexity was exponential, which is very high. In 2016, Yuan et al. \cite{Yuan2016} improved their algorithm to reduce the complexity to the polynomial level, but it is not used to solve specific problems. In 2019, Fan et al. \cite{Fan2019Q} used the basic knowledge of grayscale morphology to study gradient images, which further expands the quantum image morphology research. After that, Li et al. \cite{Li2019} made practical applications of morphological theory, such as edge detection, image enhancement.  As far as we know, morphological theory can also be well applied to image segmentation with uneven illumination. In the grayscale level image, the top-hat transform structure element moves below the grayscale level to obtain the open operation image. Then we subtract the open operation image from the original image, and the background can be removed.The bottom-hap transform structure element moves above the grayscale level to obtain a closed operation image. Then we subtract the original image from the closed operation image, and the background can be removed. Therefore, the morphological operation can well remove the uneven illumination in the image, which is beyond other methods.  However,
each pixel in the image needs  to be processed  individually, so it requires a lot of computation time   when the scale of image increase, which  leads to the real-time problem.  In order to solve the problem,  a quantum image segmentation algorithm using the grayscale morphology theory is proposed. Unlike classical methods, quantum methods can take advantage of the parallelism of quantum computing to process the quantum superposition state pixels of the input image, which greatly speeds up the classical counterpart. 

In a nutshell, the main contributions of this work are as follows:

\begin{itemize}
    \item A  quantum image segmentation algorithm based on grayscale morphology  is firstly proposed, which can use quantum mechanism to simultaneously perform morphological operations on all pixels in a grayscale image and quickly segment the image with uneven lighting into a binary image.
    
    \item Several specific quantum circuit units, including dilation, erosion, bottom hat transformation, top hat transformation, etc., are designed in detail by using fewer qubits and quantum gates. And then based on these units,  the complete quantum  circuits are designed to segment the NEQR images.
    
    \item  We verify the superiority and feasibility of our proposed algorithm by analyzing the circuit complexity and performing  simulation experiments on IBM Q, respectively.
\end{itemize}

The rest of this  paper is organized as follows. Section \ref{Section 2} introduces the principle of the NEQR representation model and the classical grayscale morphology. In Sect. \ref{Section 3}, some basic quantum operation modules are introduced, then, a series of quantum circuit units  are designed and some relevant quantum states equations are given. Section \ref{Section 4} analyzes the circuit complexity of our algorithm and the experiment is shown in Sect. \ref{Section 5}. Finally,  the conclusion is drawn in Sect. \ref{Section 6}.

\section{Related work}\label{Section 2}
\subsection{NEQR}
The NEQR model stores the grayscale information (C) and positional information (Y and X) of a classical image in three qubit sequences respectively, and entangles the position and grayscale information to form a complete quantum image. For a classical image with a size of $2^n \times2^n$ and a grayscale range of $[0,2^q-1]$, $2n$ qubits are required to store the position information of the Y-axis and X-axis. Then, $q$ qubits are required to store the grayscale value information. The expression is shown in Eq. \ref{eq1} \cite{Zhang2013}.

\begin{equation}\label{eq1}
\begin{aligned}
   \lvert {\rm{I}} \rangle  &= \frac{1}{{{2^n}}}\sum\limits_{Y = 0}^{{2^n} - 1} {\sum\limits_{X = 0}^{{2^n} - 1} {\lvert {{C_{YX}}} \rangle  \otimes \lvert Y \rangle \lvert X \rangle } }  \\
   &= \frac{1}{{{2^n}}}\sum\limits_{YX = 0}^{{2^{2n}} - 1} {\mathop  \otimes \limits_{k = 0}^{q - 1}\lvert {C^K_{YX}} \rangle \mathop  \otimes \limits_{}^{} \lvert {YX} \rangle  } 
   \end{aligned}
\end{equation}
where  $\lvert {{C_{YX}}} \rangle  = \lvert {C_{YX}^{q - 1},C_{YX}^{q - 2},{ \cdots ^{}}C_{YX}^{1}C_{YX}^{0}} \rangle$ represents the quantum image gray-scale values, $C_{YX}^k \in \left\{ {0,1} \right\}$, $k = q - 1,q - 2, \cdots ,0$. $  \lvert {YX} \rangle  = \lvert Y \rangle \lvert X\rangle  = \lvert {{Y_{n - 1}},{Y_{n - 2}}, \cdots {Y_0}}\rangle \lvert {{X_{n - 1}},{X_{n - 2}}, \cdots {X_0}} \rangle $ represents the 
position of the pixel in a quantum image, ${Y_t},{X_t} \in \left\{ {0,1} \right\}$.

Figure \ref{Fig1} shows an example of a grayscale image of size $2\times2$, and the corresponding NEQR expression of which is given as follows.
\begin{equation}
\begin{aligned}
\lvert I \rangle & = \frac{1}{2}\left( {\lvert 0 \rangle \lvert{00} \rangle  + \lvert {100} \rangle \lvert {01} \rangle  + \lvert {200} \rangle \lvert {10} \rangle  + \lvert{255} \rangle \lvert {11} \rangle } \right)
\\
&=\frac{1}{2}\left( \begin{array}{l}
\lvert {00000000} \rangle \lvert{00} \rangle  + \lvert {01100100} \rangle \lvert {01} \rangle \\
 + \lvert {11001000}\rangle \lvert{10} \rangle  + \lvert {11111111} \rangle \lvert {11} \rangle 
\end{array} \right)
\end{aligned}
\end{equation}

\Figure[t!](topskip=0pt, botskip=0pt, midskip=0pt){Fig1}
{An example of $2\times2$ image.\label{Fig1}}

\subsection{Classical grayscale morphology}

\subsubsection{Grayscale dilation and erosion}
Grayscale morphology is extended from binary morphology, which allows grayscale images to be processed with morphological operations as well. Grayscale morphology also includes two basic operations: dilation and erosion, but the original  “AND" and  “OR" operations are turned into a problem of finding local extremes.  The formula expression is as follows.
\begin{equation}
    [F \oplus {B_N}](x,y) = \mathop {\max }\limits_{(s,t) \in {D_{{B_N}}}} \left\{ {F(x - s,y - t) + {B_N}(s,t)} \right\}
\end{equation}
where $ {{B_N}}$ represents the origin symmetric transformation of the non-flat structure element $\mathop {{B_N}}\limits^ \wedge $, and ${D_{{B_N}}}$ represents the region of $ {{B_N}}$. $x$ and $y$ are incremented so that the midpoint of $ {{B_N}}$ can access every pixel in $F$. However, non-flat structure element will increase the amount of calculation and affect the result image. Therefore, in practical use, flat structure elements are often used to perform the grayscale 
dilation operation. At this time,  $ {{B_N}=0}$, and the dilation operation is simplified as follows.
\begin{equation}
    [F \oplus B](x,y) = \mathop {\max }\limits_{(s,t) \in {D_B}} \left\{ {F(x - s,y - t)} \right\}
\end{equation}
where $F$ represents the original image, $B$ represents the structure element, and $D_B$ represents the region of $B$. In this case, the dilation operation is to find the maximum value of the overlapping area of $F$ and $B$, and assign it to the center pixel of the overlapping area. 

Similar to the dilation operation, the erosion operation can be expressed as follows.

\begin{equation}
    [F\Theta {B_N}](x,y) = \mathop {\min }\limits_{(s,t) \in {D_{{B_N}}}} \left\{ {F(x + s,y + t) - {B_N}(s,t)} \right\}
\end{equation}

\begin{figure}
    \centering
    \subfigure[]{\includegraphics[width=2.5cm]{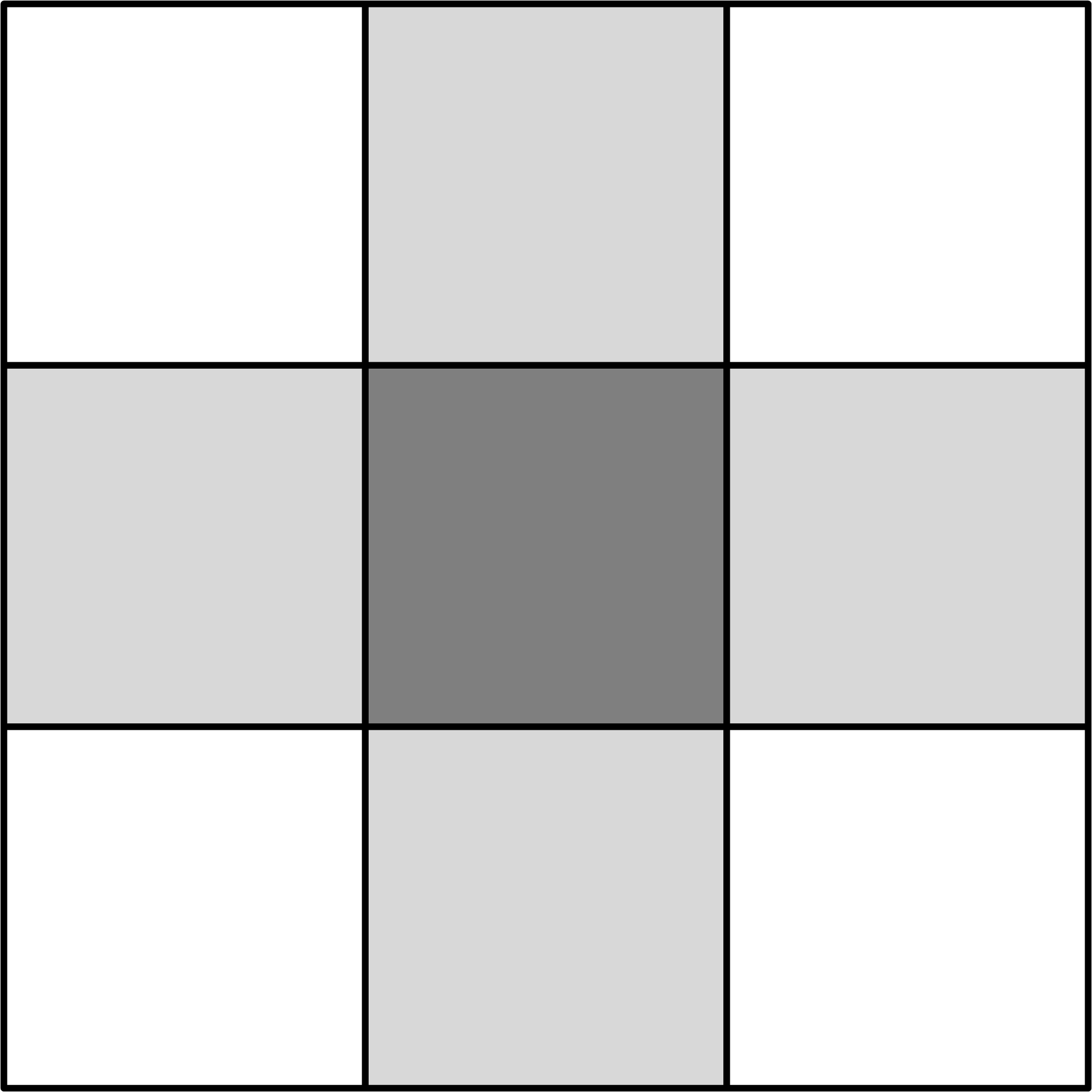}}
    \subfigure[]{\includegraphics[width=2.5cm]{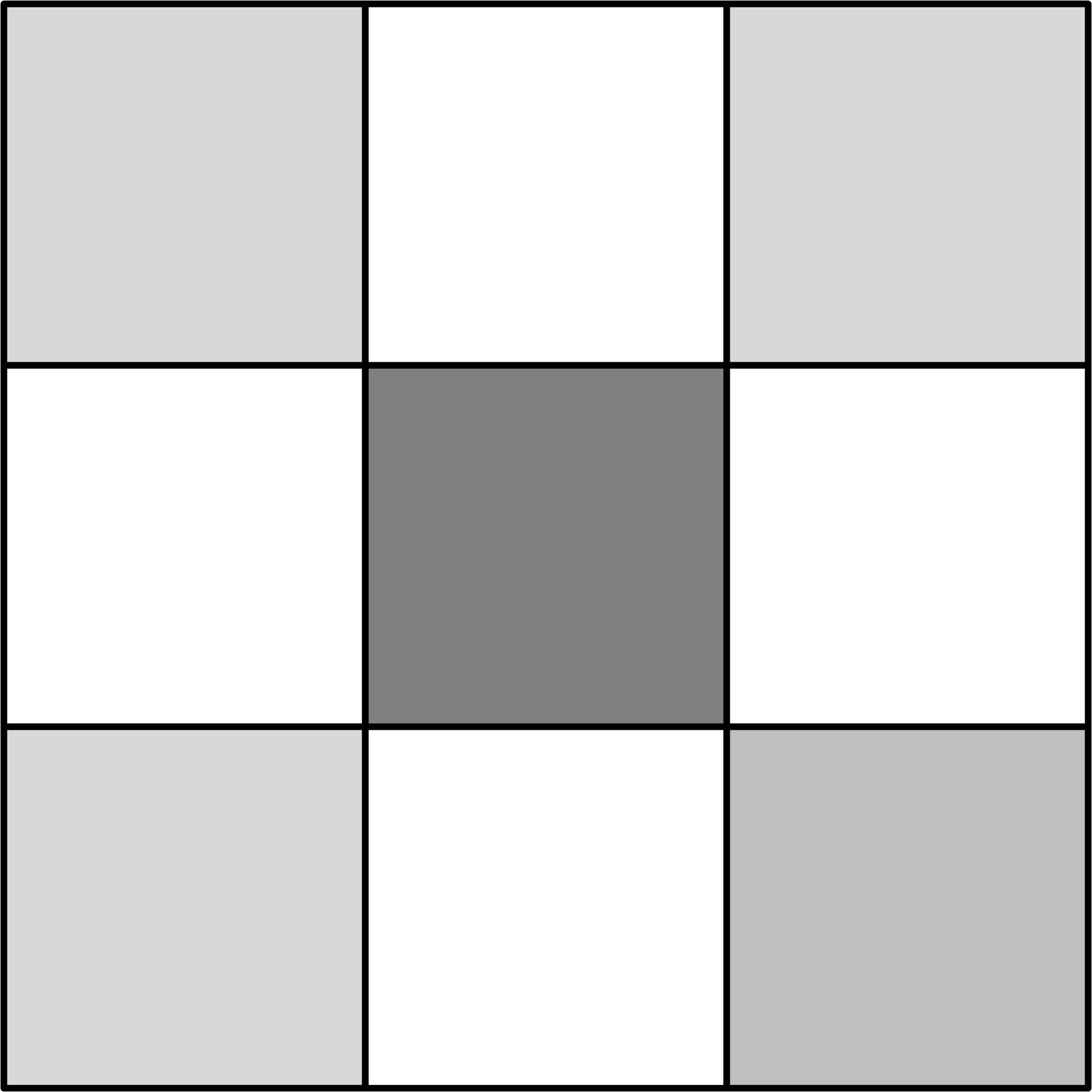}}
    \subfigure[]{\includegraphics[width=2.5cm]{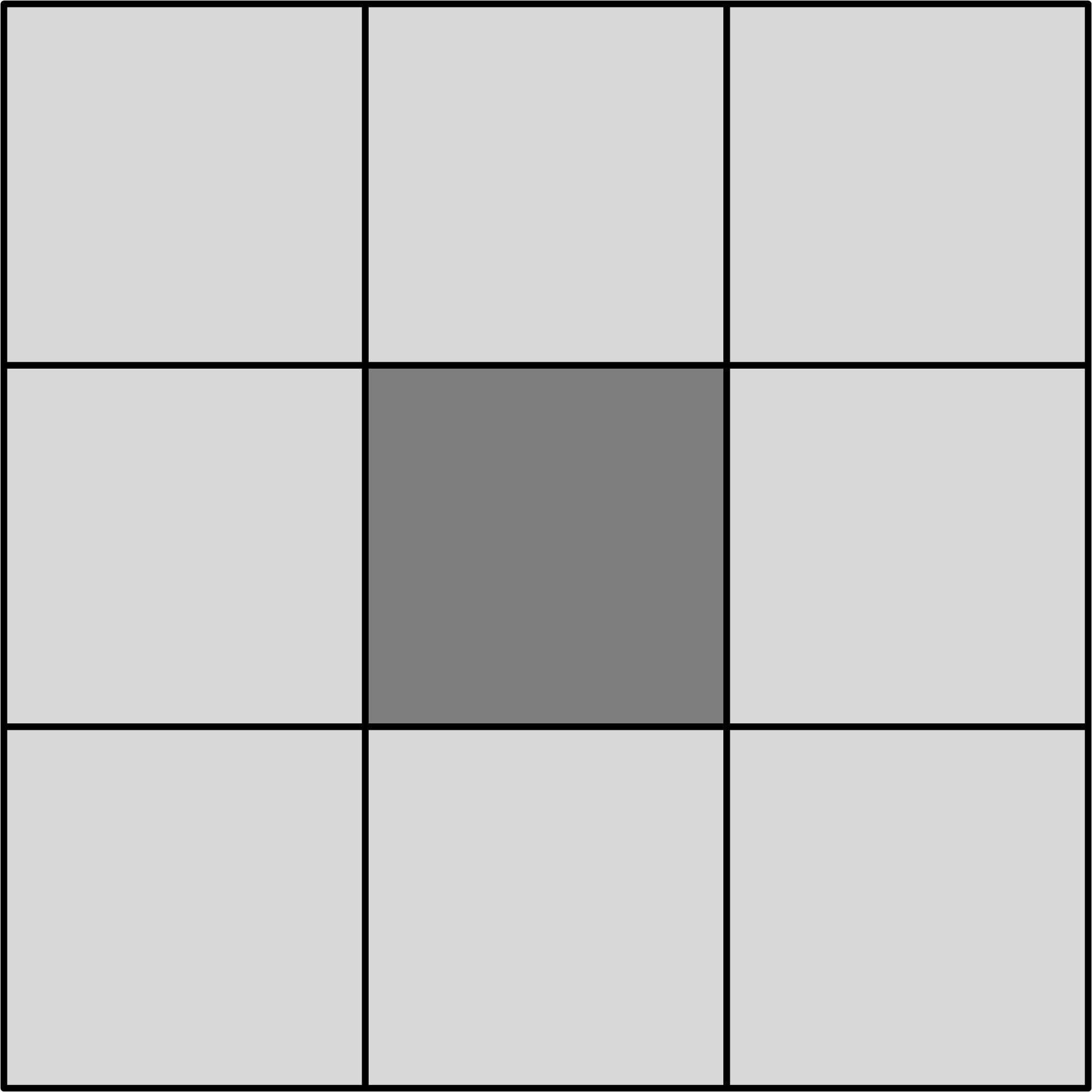}}
    \caption{The schematic diagram of the structuring windows.}
    \label{Fig2}
\end{figure}

Among them, $ {{B_N}}$ represents the non-flat structure element, and ${D_{{B_N}}}$ represents the region of $ {{B_N}}$. Also in practical use, flat structure elements are often used for grayscale erosion operation. In this case, $ {{B_N}}=0$, and the erosion operation is simplified as follows.
\begin{equation}
    [F\Theta B](x,y) = \mathop {\min }\limits_{(s,t) \in {D_B}} \left\{ {F(x + s,y + t)} \right\}
\end{equation}
where $F$ represents the original image, $B$ represents the structure element, and $D_B$ represents the region of $B$. In this case, the erosion operation is to find the minimum value of the overlapping area of $F$ and $B$, and assign it to the center pixel of the overlapping area. We use different structure elements for different images to get better results, and the commonly used structuring elements are shown in Fig. \ref{Fig2}.

\subsubsection{Bottom hat transformation and top hat transformation}
\begin{figure}
    \centering
    \includegraphics[width=8cm]{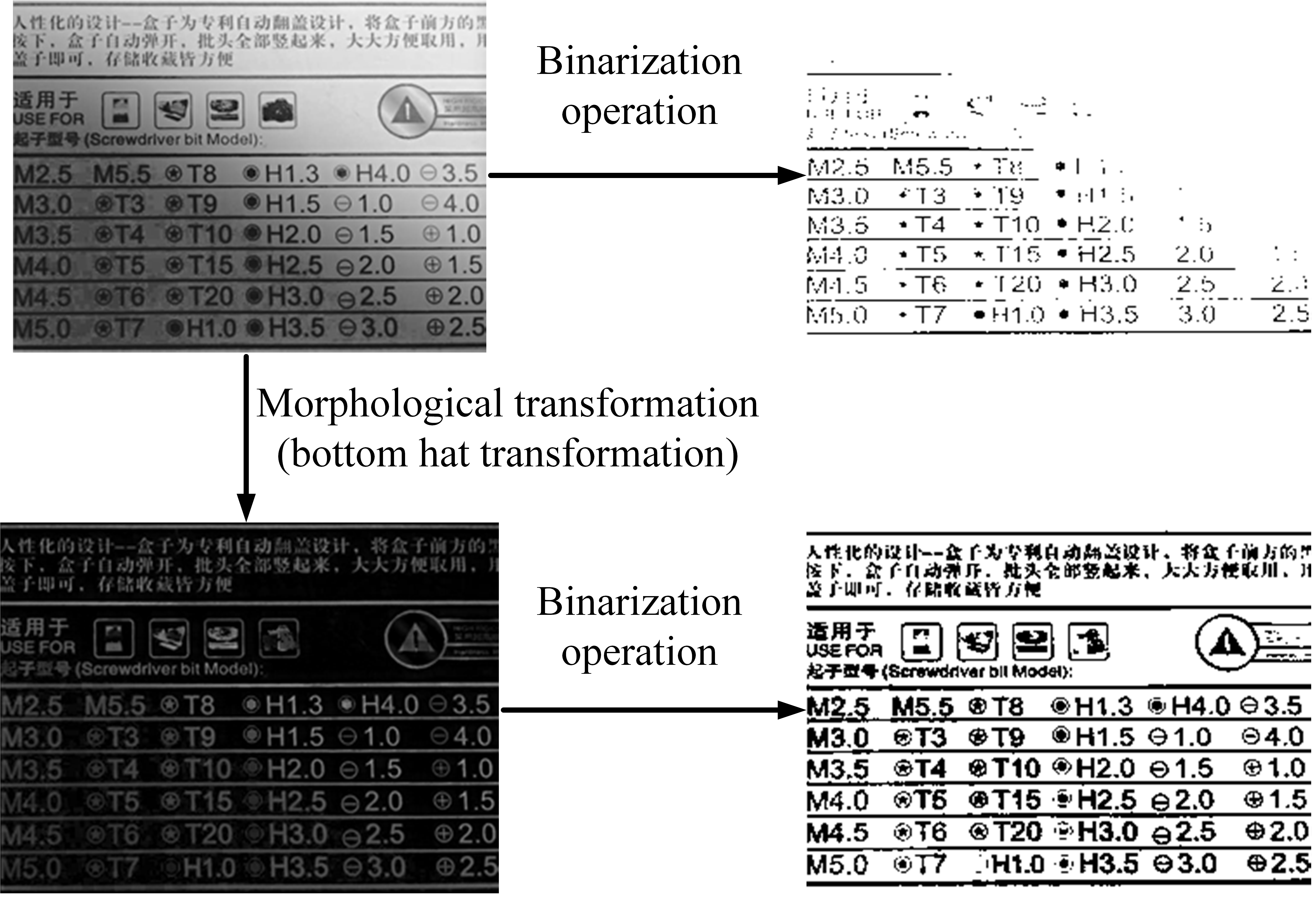}
    \caption{Segmentation of unevenly illuminated image.}
    \label{Fig3}
\end{figure}

In the practical application of image segmentation, there will be uneven light in the image, which will make it difficult to find the objects and bring errors to the segmentation. Morphological bottom hat and top hat transformation can solve this problem very well. As shown in Fig. \ref{Fig3}, it is an image with uneven background illumination. When the image is directly segmented, part of the text will be lost; when the bottom hat transformation is used, the background with uneven illumination in the image is deleted. So that the  complete text information can be obtained by performing the binaryzation operation.

The bottom hat transformation of the grayscale image $F$ is defined as the result of the image closing operation minus the original image.  Specifically, the image is first dilated and then eroded, and finally we subtract the original image from the processed image. The bottom hat transformation can get the dark area in the original image, so it is also called the black bottom hat transformation. The definitions are as follows.
\begin{equation}
    {B_{hat}}(F) = F \bullet B - F = [(F \oplus B)\Theta B] - F
\end{equation}

The top hat transformation of grayscale image $F$ is defined as the original image minus the result image of the opening operation. Firstly,  the image is eroded, then it is dilated, and finally we subtract the previously obtained image from the original image. The top hat transformation can get the bright area in the original image, so it is also called the white top hat transformation, and it can  be expressed
as follows.
\begin{equation}
    {T_{hat}}(F) = F - (F \circ B) = F - [(F\Theta B) \oplus B]
\end{equation}

\section{Quantum image segmentation algorithm based on grayscale morphology}\label{Section 3}

\subsection{Quantum operations}

\begin{enumerate}[(1)]

\item \textbf{Quantum comparator} 

The quantum comparator can compare the numerical magnitudes of two binary qubit sequences. It takes two qubit sequences $\lvert A \rangle$ and $\lvert B \rangle$ as input and outputs the two sequences  and their comparison result $y$. Inspired by the quantum  bit string comparator (QBSC) \cite{Oliveira2007}, our comparator also compares bit-by-bit. The comparison results for each bit are processed using three auxiliary qubits. If $A\ge B$, then $y=0$; if $A<B$, then $y=1$. The detailed quantum circuit diagram is shown in Fig. \ref{Fig4}. Compared with the existing quantum comparators, our comparators use fewer quantum gates and fewer qubits, as shown in Tab. \ref{tab1}, which is very necessary in this NISQ era.

For further sorting operation, we use a CSWAP gate at the output of the comparator to control the output order of $A$ and $B$, as shown in the Fig. \ref{Fig5}, QCS means ${A'} \ge {B'}$, QCL means ${A'}<{B'}$.

\begin{table}
\caption{ Comparison of different quantum comparators.}\label{tab1}
\begin{tabular}{@{}ccc@{}}
\toprule
{ \textit{Quantum comparators}} & { \textit{Auxiliary qubits}} & { Quantum cost} \\ \midrule
{ \textit{QBSC}} \cite{Oliveira2007}           & $3n-1$ & { $30n-15$} \\
{ \textit{QC}} \cite{Yuan2020}            & { 5}    & { $28n-15$} \\
{ \textit{Our comparator}} & {3}    & { $18n-3$}  \\ \bottomrule
\end{tabular}
\end{table}

\Figure[t!](topskip=0pt, botskip=0pt, midskip=0pt)[width=15 cm]{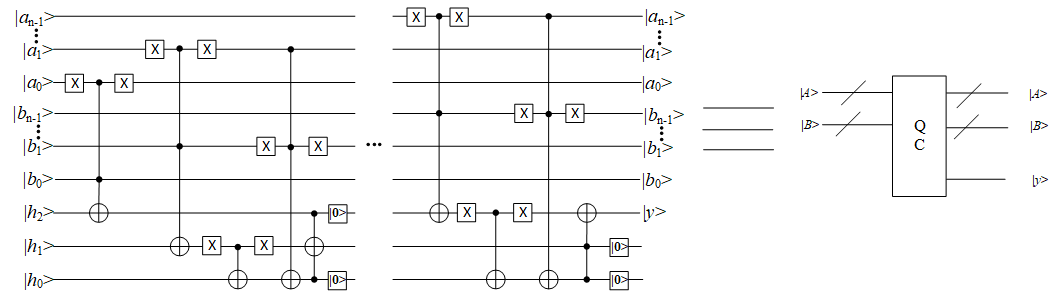}
{Implementation circuit of the quantum comparator.\label{Fig4}}

\begin{figure*}
    \centering
    \subfigure[QCS]{\includegraphics[width=5.6cm]{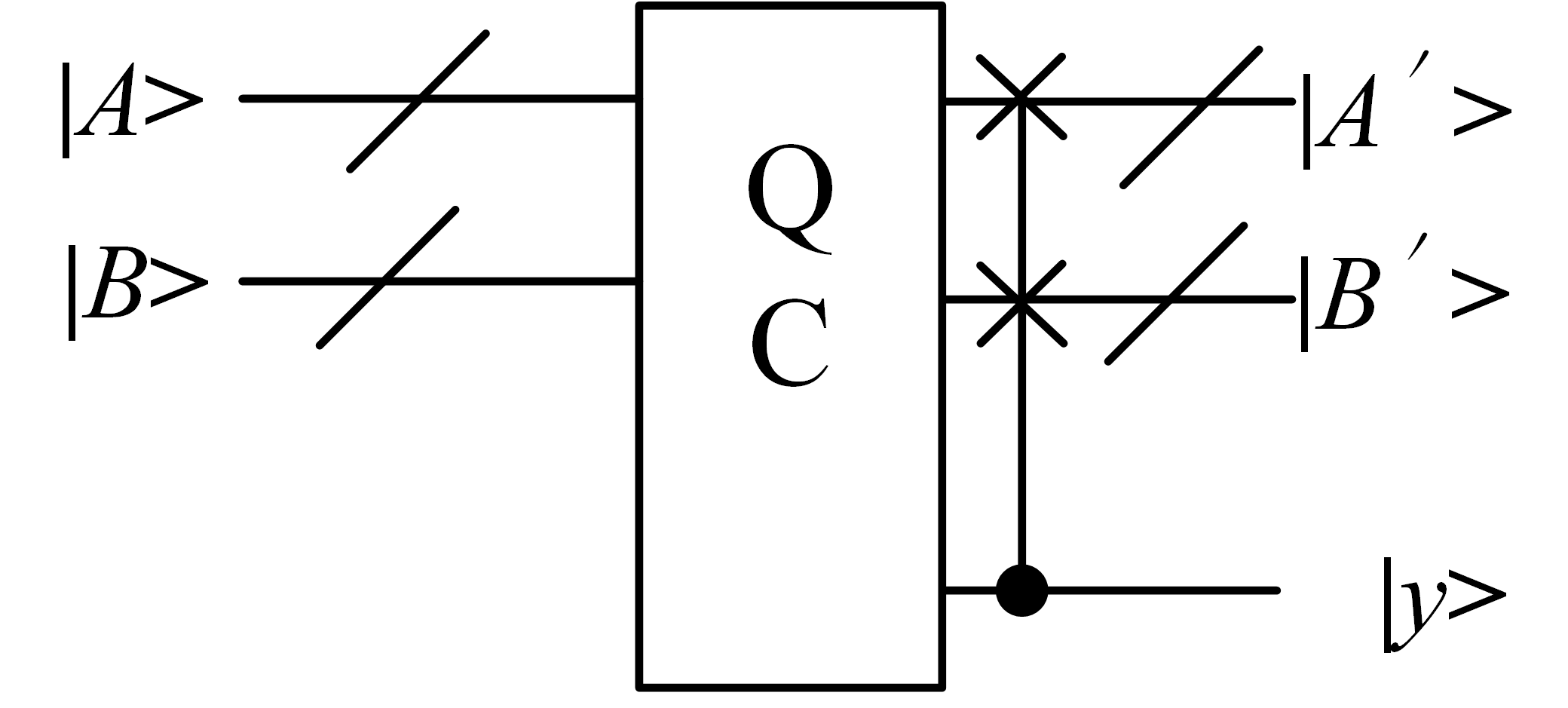}}
     \subfigure[QCL]{\includegraphics[width=6cm]{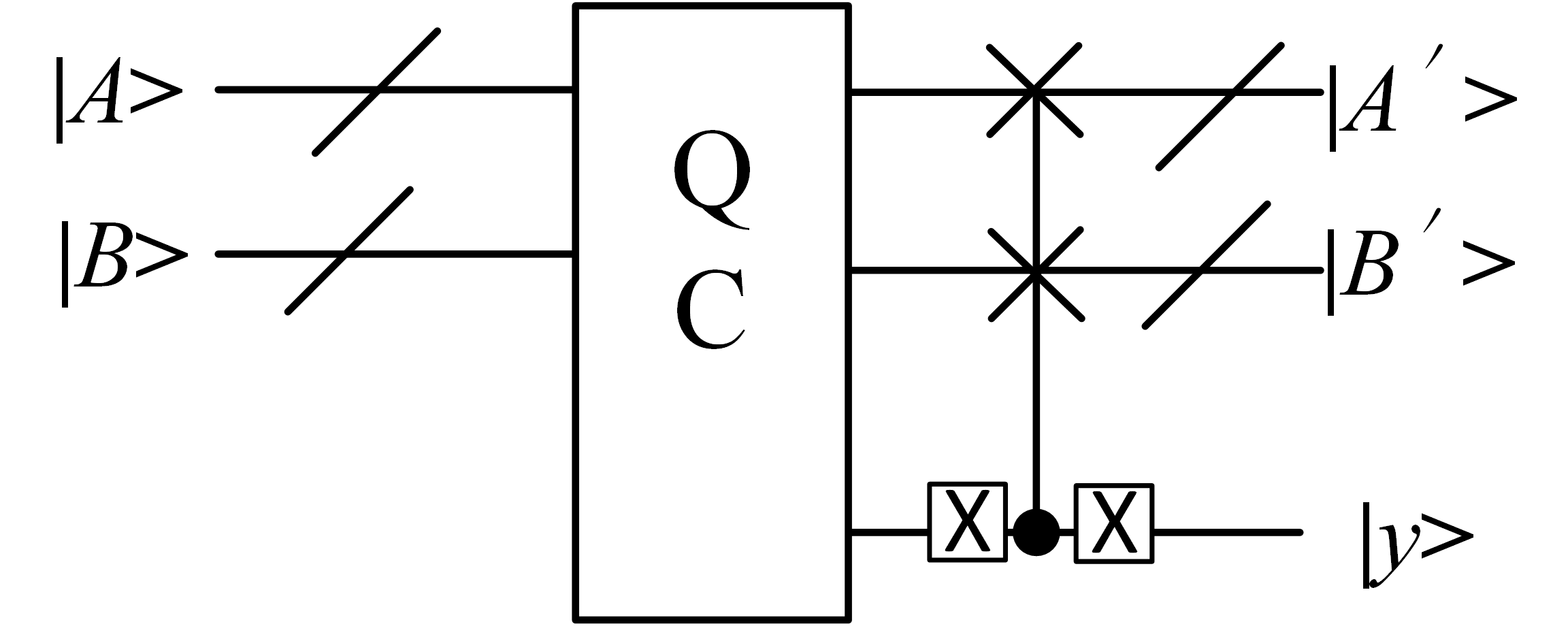}}
    \caption{Implementation circuit of the QCS and QCL.}
    \label{Fig5}
\end{figure*}

\item \textbf{Quantum subtractor operation}

The quantum subtractor can realize the subtraction operation of two binary qubit sequences $A$ and $B$. It takes $A$ and $B$ as inputs and outputs the result of $A-B$. By performing a bit-by-bit subtraction operation on ${a_0}-{b_0}, {a_1}-{b_1},...,{a_{n-1}}-{b_{n-1}}$, and retaining the borrow information for the next bit, the subtract  operation can be completed using 2 auxiliary qubits. The specific implementation circuit is shown in Fig. \ref{Fig6}.

\Figure[t!](topskip=0pt, botskip=0pt, midskip=0pt)[width=15 cm]{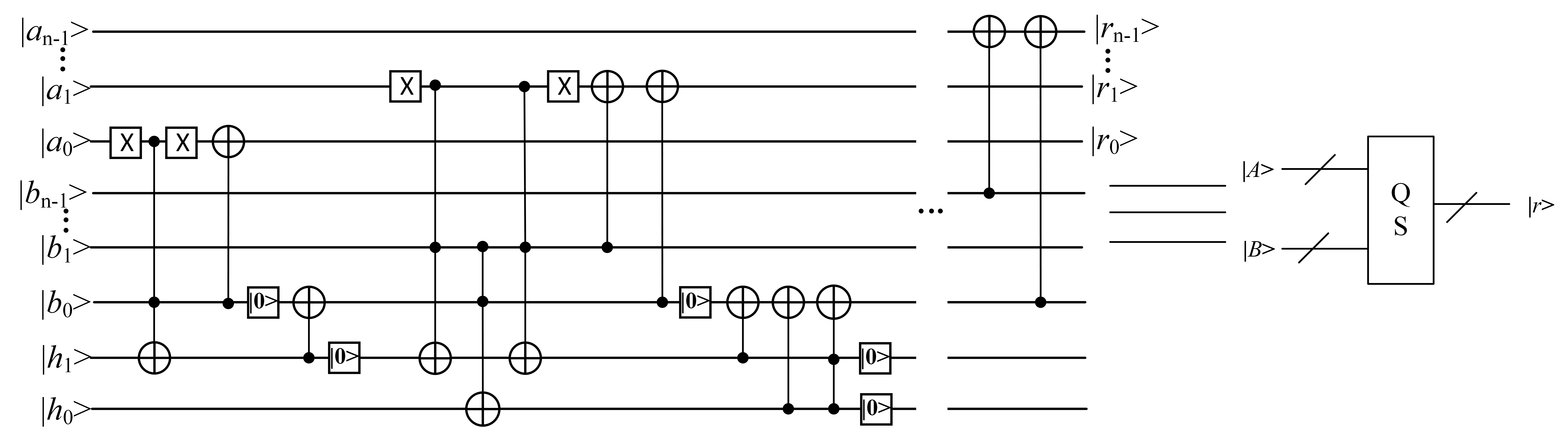}
{Implementation circuit of the quantum subtractor.\label{Fig6}}

\item  \textbf{Cycle shift transformation}
  
  The quantum image cyclic shift transformation $(CT)$ operation \cite{Le2011S}  can shift the positions of all pixels in an image, so that each pixel in the image can obtain its neighbor pixels at the same time. For example, if we move the image up by one unit $(CT_{y+})$, then the pixels in the image change from $S(X,Y)$ to $S(X,Y+1)$. The unitary operation for cyclic shift of a $2^n\times2^n$ NEQR image is shown below.
  \begin{equation}
\mathrm{CT}(X \pm)\lvert I\rangle=\frac{1}{2^{n}} \sum_{Y=0}^{2^{n}-1} \sum_{X=0}^{2^{n}-1}\lvert C T_{Y X^{\prime}} \rangle \lvert Y\rangle \lvert (X \pm 1) \bmod 2^{n}\rangle 
\end{equation}
\begin{equation}
\mathrm{CT}(Y \pm)\lvert I\rangle=\frac{1}{2^{n}} \sum_{Y=0}^{2^{n}-1} \sum_{X=0}^{2^{n}-1}\lvert C T_{Y^{\prime} X} \rangle \lvert (Y \pm 1) \bmod 2^{n}\rangle \lvert X \rangle 
\end{equation}
where $X^{\prime}=(X \mp  1) \bmod 2^{n}$, $Y^{\prime}=(Y \mp  1) \bmod 2^{n}$, $\mathrm{CT}_{(X+)} \, \&\, \mathrm{CT}_{(Y+)}=\left[ {\begin{array}{*{20}{c}}
0&1\\
{I_2^n - 1}&0
\end{array}} \right]$, $\mathrm{CT}_{(X-)} \,\&\, \mathrm{CT}_{(Y-)}=\left[ {\begin{array}{*{20}{c}}
0&{I_2^n - 1}\\
1&0
\end{array}} \right]$. Figure \ref{Fig7} shows a schematic diagram of the X-axis  cyclic shift transformation.
\begin{figure*}
    \centering
 \subfigure[]{\includegraphics[width=7cm]{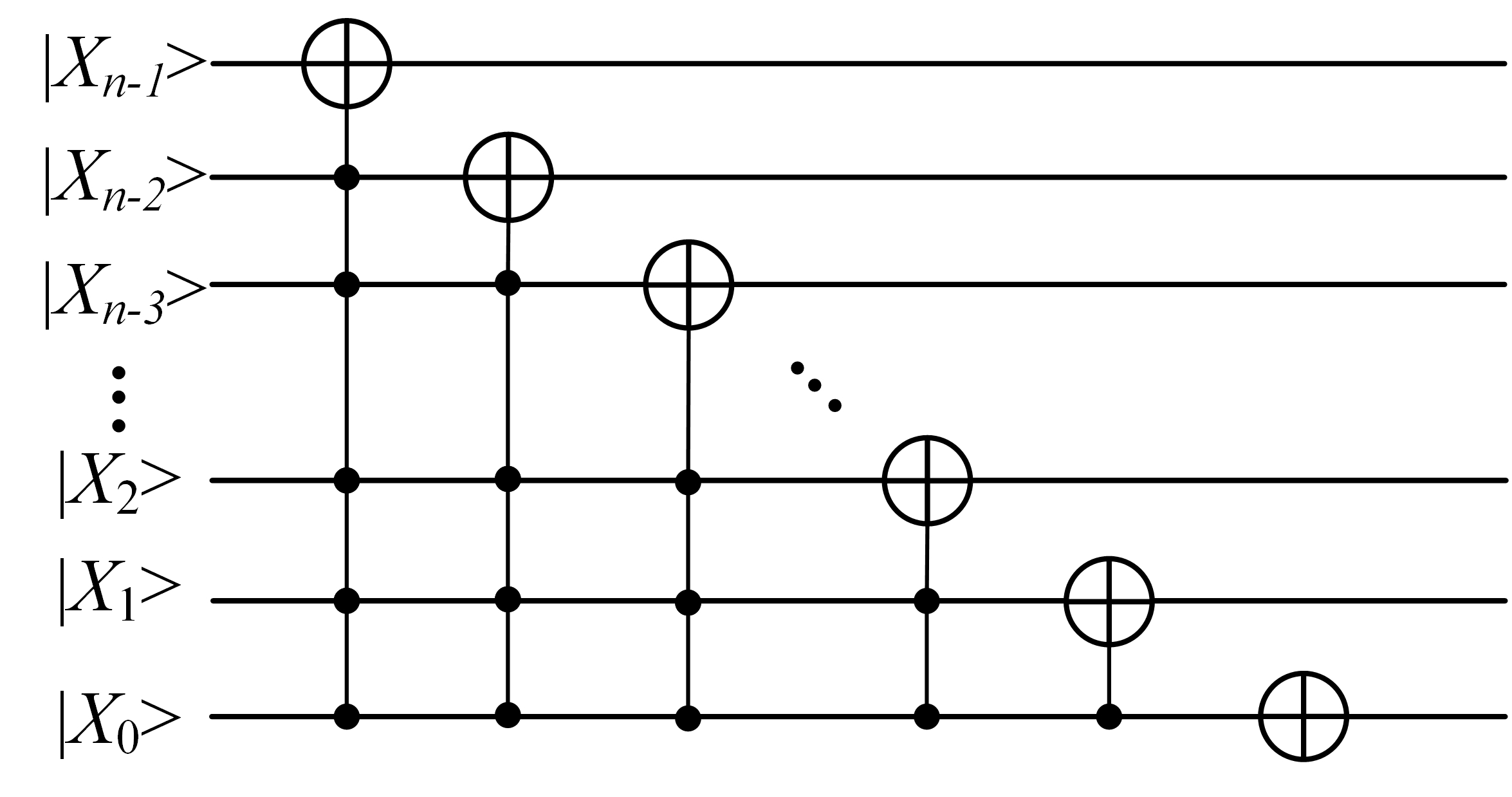}}
 \subfigure[]{\includegraphics[width=7cm]{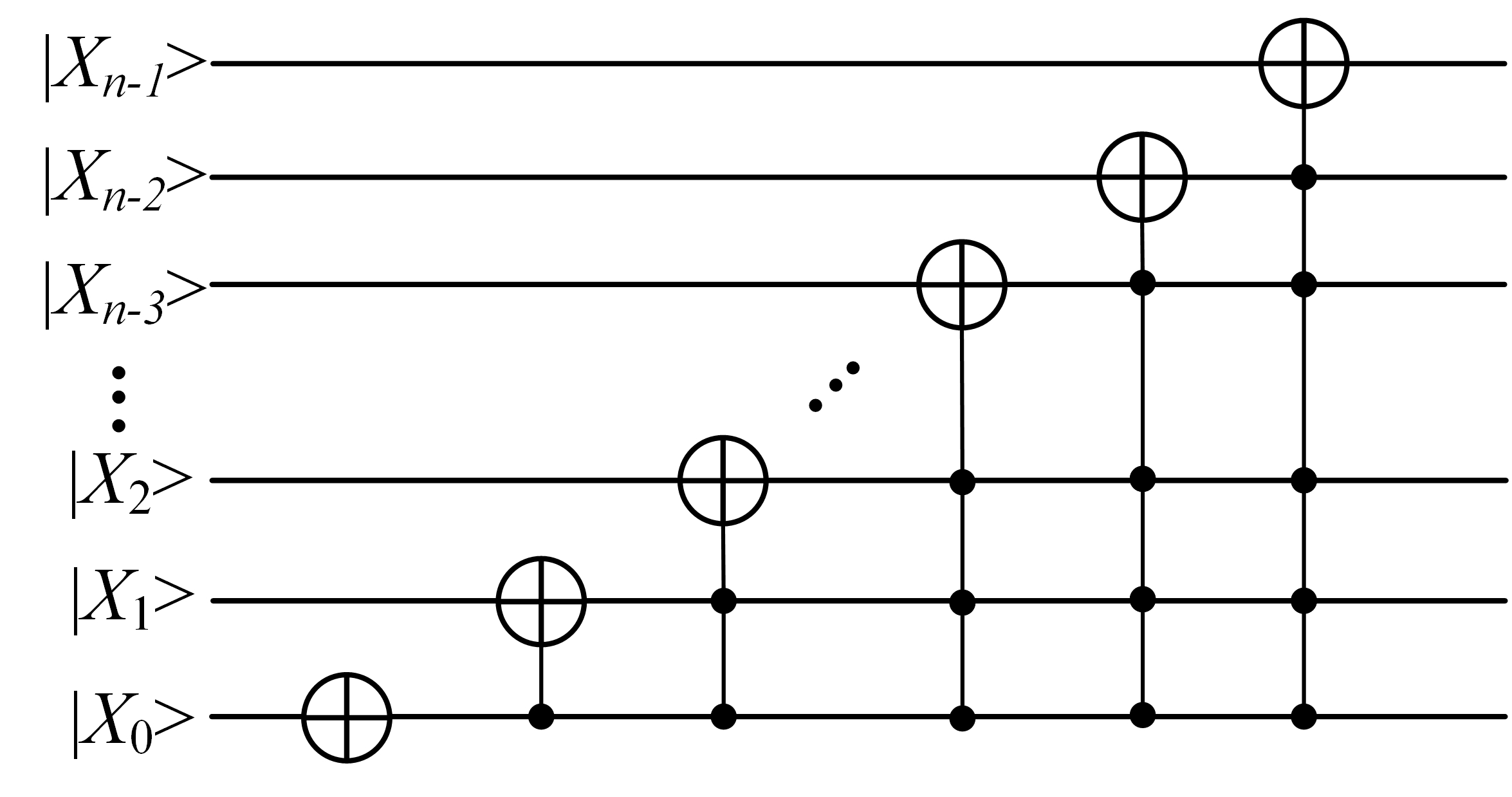}}
    \caption{The schematic diagram of the X-axis cyclic shift transformation.}
    \label{Fig7}
\end{figure*}

\item \textbf{Quantum copy operation}

In order to copy the grayscale value of the image, we need to use the Copy operation \cite{Iliyasu2013} to complete. It is implemented with CNOT gates and auxiliary qubits. The circuit is shown in Fig. \ref{Fig8}, $\lvert x\rangle$ represents the original value and $\lvert 0\rangle$ is used to store the copied value.
\begin{figure*}
    \centering
 {\includegraphics[width=11cm]{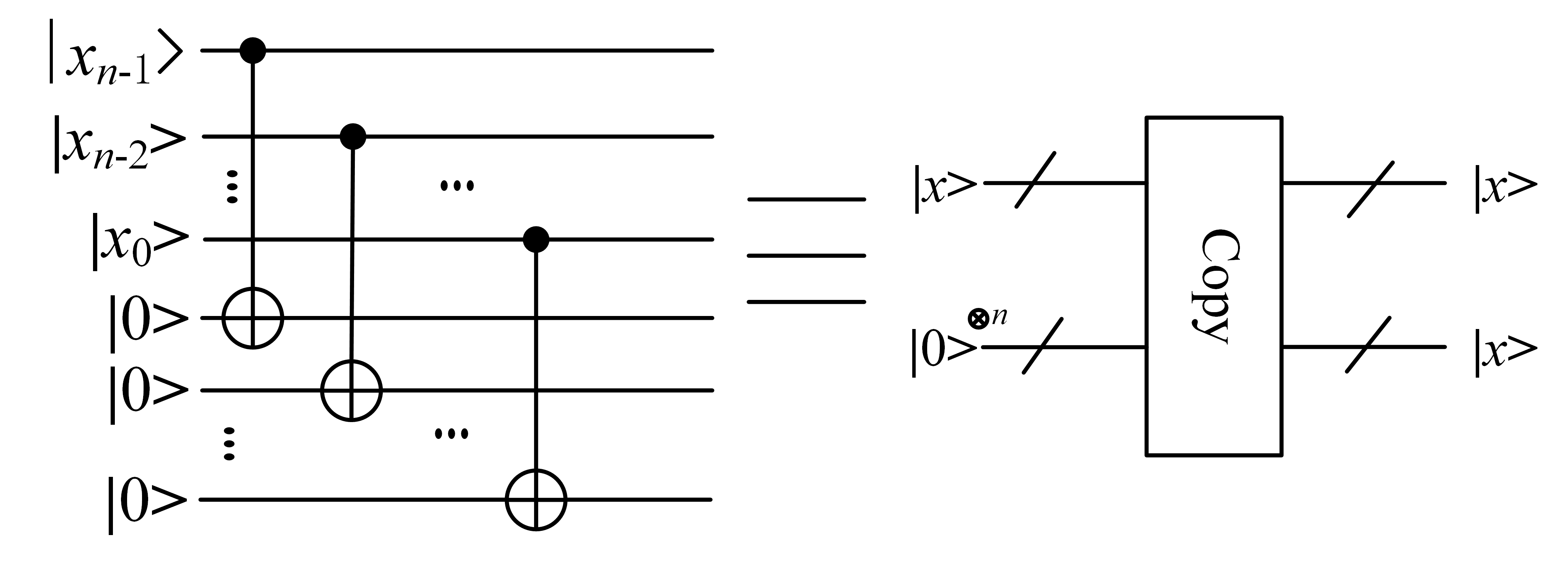}}
    \caption{The quantum circuit realization of Copy operation.}
    \label{Fig8}
\end{figure*}

\end{enumerate}

 \subsection{The quantum image segmentation algorithm}
 In order to achieve the segmentation of images with uneven illumination, our proposed quantum image segmentation algorithm can be described as Algorithm \ref{algo1}.
 
\begin{figure*}
    \centering
 \includegraphics[width=14cm]{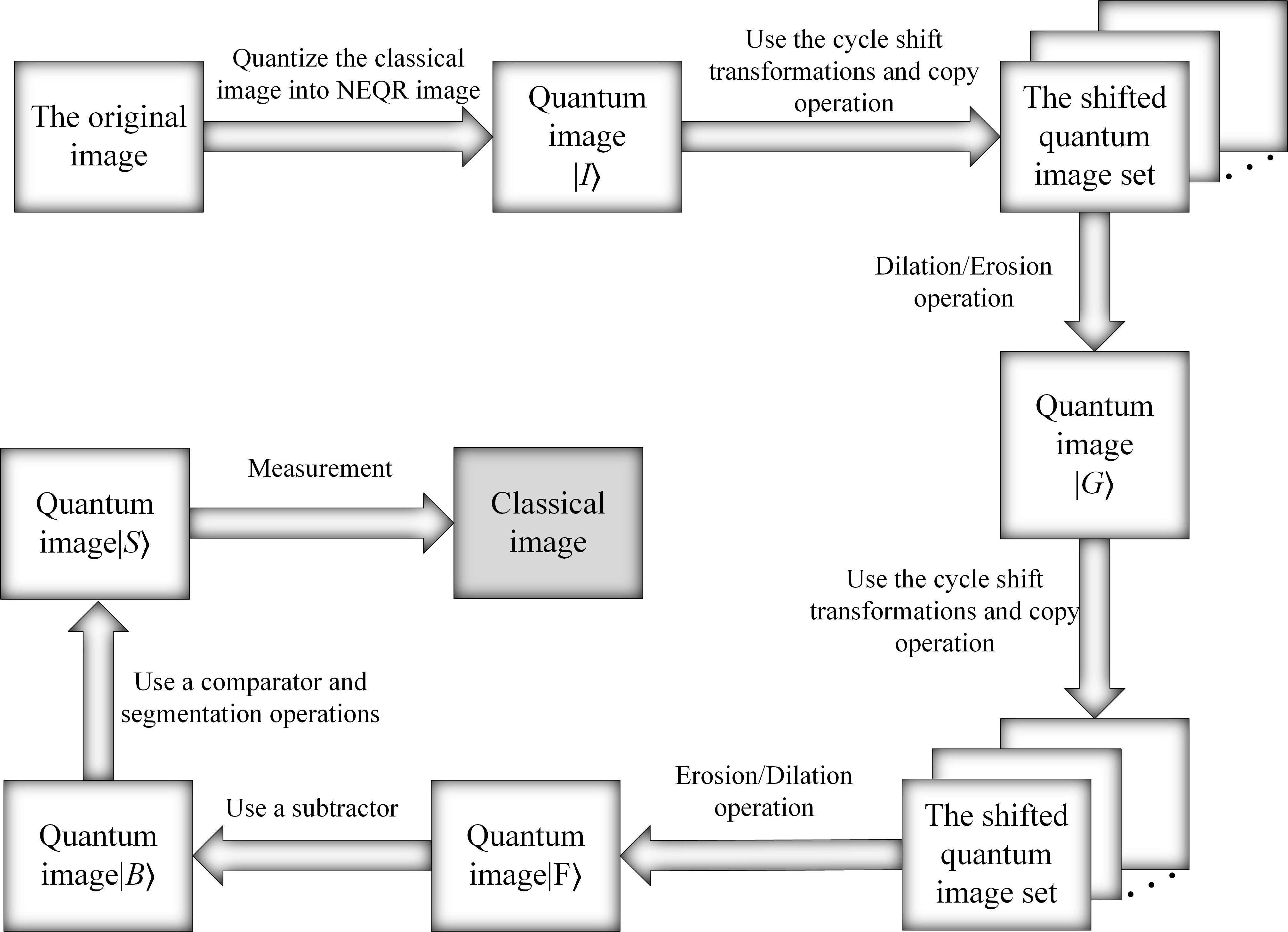}
    \caption{The workflow of our proposed algorithm.}
    \label{Fig9}
\end{figure*}

In the second step, we use the cyclic shift operation to shift the original image $\lvert I\rangle$ one unit in four directions: up, down, left, and right. Each translation requires the CT operation on the  position qubits of $\lvert I\rangle$. Then we prepare the image set according to the $\lvert I\rangle$ preparation method. After the preparation, we use CT operation to restore the position to the position information of  $\lvert I\rangle$. In this way, five images can be obtained after four CT operation, and their position information shares the  position qubits of $\lvert I\rangle$. For an image containing bright backgrounds and dark targets, we need to perform a bottom-hat transformation on it. That is, we first perform the dilation operation on the original image to get the image $\lvert G\rangle$, then we cycle shift $\lvert G\rangle$ to get an image set, then we perform the erosion operation on $\lvert G\rangle$ to get the image $\lvert F\rangle$, and finally we subtract the original image $\lvert I\rangle$ from image $\lvert F\rangle$. On the contrary, for the image containing dark background and bright target, we need to perform top-hat transformation on it. In other words, the original image is first eroded to get the image $\lvert G\rangle$, then $\lvert G\rangle$ is cyclically shifted to get the image set, then the dilation operation is performed on $\lvert G\rangle$ to get the image $\lvert F\rangle$, and finally we subtract $\lvert F\rangle$ from the original image $\lvert I\rangle$. After the morphological operation, the uneven light in the image can be deleted. At this time, the object can be segmented by binary operation. The complete process is shown in Fig. \ref{Fig9}.

  \begin{algorithm}
\caption{The quantum image segmentation algorithm.}\label{algo1}
\begin{algorithmic}[1]
\STATE Preparing the quantum image $\lvert I\rangle$ based on NEQR model.
\STATE Using the cyclic shift operation to prepare a quantum image set of $\lvert I\rangle$.
\STATE Using the quantum image set, the NEQR image  $\lvert I\rangle$ is morphologically processed by dilation/erosion operation, and the result image is named  $\lvert G\rangle$.
\STATE Using the cyclic shift operation again to prepare the quantum image set of $\lvert G\rangle$.
\STATE Using the quantum image set of $\lvert G\rangle$, the NEQR image $\lvert G\rangle$ is processed by erosion/dilation operation, and the result image is named $\lvert F\rangle$.
\STATE  Using the subtractor to make $\lvert F\rangle-\lvert I\rangle$ / $\lvert I\rangle-\lvert F\rangle$, then the top hat transform/bottom hat transform result $\lvert B\rangle$ is obtained.
\STATE Using the comparator and the segmentation operation to segment $\lvert B\rangle$, and finally the binary image $\lvert S\rangle$ is obtained.
\STATE The quantum measurement operation is performed to retrieve the classical image, and the algorithm ends.)
\end{algorithmic}
\end{algorithm}
 
 \subsection{Quantum circuit implementation of the algorithm}

\begin{enumerate}[(1)]

\item \textbf{Quantum image preparation } 

According to the NEQR model, $2n$ qubits are needed to store the position information, and $q$ qubits are needed to store the grayscale value information. In addition, additional $4q$ qubits are also required to store the grayscale values of the 4-neighborhood pixels. Because the neighborhood window will be shuffled after grayscale morphological processing, in order to process the original image later, we still need additional $q$ qubits to store the grayscale information of the original image. The quantum state expression is as follows.

\begin{equation}
    \begin{array}{l}
{\lvert 0 \rangle ^{ \otimes 5q}} \otimes \lvert {{I_{YX}}}  \rangle  = \frac{1}{{{2^n}}}\sum\limits_{Y = 0}^{{2^n} - 1} {\sum\limits_{X = 0}^{{2^n} - 1} {{{\lvert 0  \rangle }^{ \otimes 5q}}\lvert {{C_{YX}}}  \rangle } } \lvert Y \rangle \lvert X \rangle \\
\begin{array}{*{20}{c}}
{}&{}&{\begin{array}{*{20}{c}}
&{}= 
\end{array}}
\end{array}\frac{1}{{{2^n}}}\sum\limits_{Y = 0}^{{2^n} - 1} {\sum\limits_{X = 0}^{{2^n} - 1} {{{\lvert 0 \rangle }^{ \otimes q}}} }  \cdots {\lvert 0 \rangle ^{ \otimes q}}\lvert Y  \rangle \lvert X \rangle 
\end{array}
\end{equation}

\item \textbf{Quantum image cycle shift transformation}

In order to be able to process all pixels in the image at the same time, we perform cyclic shift transformation on the NEQR image according to the structure element window, so that we can get a quantum image set, as shown in Tab. \ref{tab2}. In this way, the pixels at the same position (Y, X) in the image set are composed of the neighborhood pixels of the structure element window of the original image. In this way, when we process the pixels in the same position in the image set, we can process all the pixels in the structure element area. We take a $3\times 3$ image as an example, as shown in  Fig. \ref{Fig10}, and the images are stored in the quantum registers in the superposition state. In order to get the neighborhood pixels, we use the cyclic shift transformation to move the original image up, down, left, right, and store it in the auxiliary qubit(four other quantum registers). In this way, we can process all the neighboring pixels at once.

Furthermore, before proceeding to the next step, we need to copy the original image into auxiliary qubits, so that the original image can be manipulated again after the dilation and erosion operation. The quantum state expression of the quantum image set after cyclic shift transformation is shown below.

\begin{table*}
\caption{Specific operations for shifting the image.}\label{tab2}%
\centering
\begin{tabular}{@{}l@{}}
\hline
1. Input: the original NEQR image $\lvert {I_{YX}} \rangle$,$ \lvert {{I_{YX}}} \rangle  = \frac{1}{{{2^n}}}\sum\limits_{Y = 0}^{{2^n} - 1} {\sum\limits_{X = 0}^{{2^n} - 1} {\lvert {{C_{YX}}} \rangle } } \lvert Y \rangle \lvert X \rangle $\\  
2. Shift ${I_{YX}}$one unit upward, then$\lvert  {{I_{Y + 1X}}} \rangle  = \frac{1}{{{2^n}}}\!\sum\limits_{Y = 0}^{{2^n} - 1} \!{\sum\limits_{X = 0}^{{2^n} - 1} {\lvert  {{C_{Y + 1X}}} \rangle } } \lvert  Y \rangle \lvert  X \rangle $\\
3. Shift${I_{YX}}$one unit downward, then$ \lvert {{I_{Y - 1X}}} \rangle  =  \frac{1}{{{2^n}}}\sum\limits_{Y = 0}^{{2^n} - 1} {\sum\limits_{X = 0}^{{2^n} - 1} {\lvert  {{C_{Y - 1X}}} \rangle } } \lvert  Y \rangle \lvert  X \rangle $\\
4. Shift${I_{YX}}$ one unit leftward,  then then$\lvert {{I_{YX + 1}}} \rangle  =  \frac{1}{{{2^n}}}\sum\limits_{Y = 0}^{{2^n} - 1} {\sum\limits_{X = 0}^{{2^n} - 1} {\lvert {{C_{YX + 1}}} \rangle } } \lvert Y \rangle \lvert X \rangle $\\

5. Shift${I_{YX}}$one unit rightward,  then $\lvert {{I_{YX - 1}}} \rangle  =  \frac{1}{{{2^n}}}\sum\limits_{Y = 0}^{{2^n} - 1} {\sum\limits_{X = 0}^{{2^n} - 1} {\lvert {{C_{YX - 1}}} \rangle } } \lvert Y \rangle \lvert X \rangle$\\
\hline
\end{tabular}
\end{table*}

\begin{equation}
\begin{aligned}
    \frac{1}{{{2^n}}}\sum\limits_{Y = 0}^{{2^n} - 1} {\sum\limits_{X = 0}^{{2^n} - 1} {\lvert {{C_{Y + 1X}}} \rangle } }  \otimes \lvert {{C_{YX + 1}}}  \rangle  \otimes \lvert {{C_{Y - 1X}}}  \rangle \\
    \otimes \lvert {{C_{YX - 1}}} \rangle  \otimes \lvert {{C_{YX}}} \rangle  \otimes \lvert {{C_{YX}}} \rangle  \otimes \lvert Y \rangle \lvert X \rangle 
    \end{aligned}
\end{equation}

\begin{figure*}
    \centering
  \includegraphics[width=15cm]{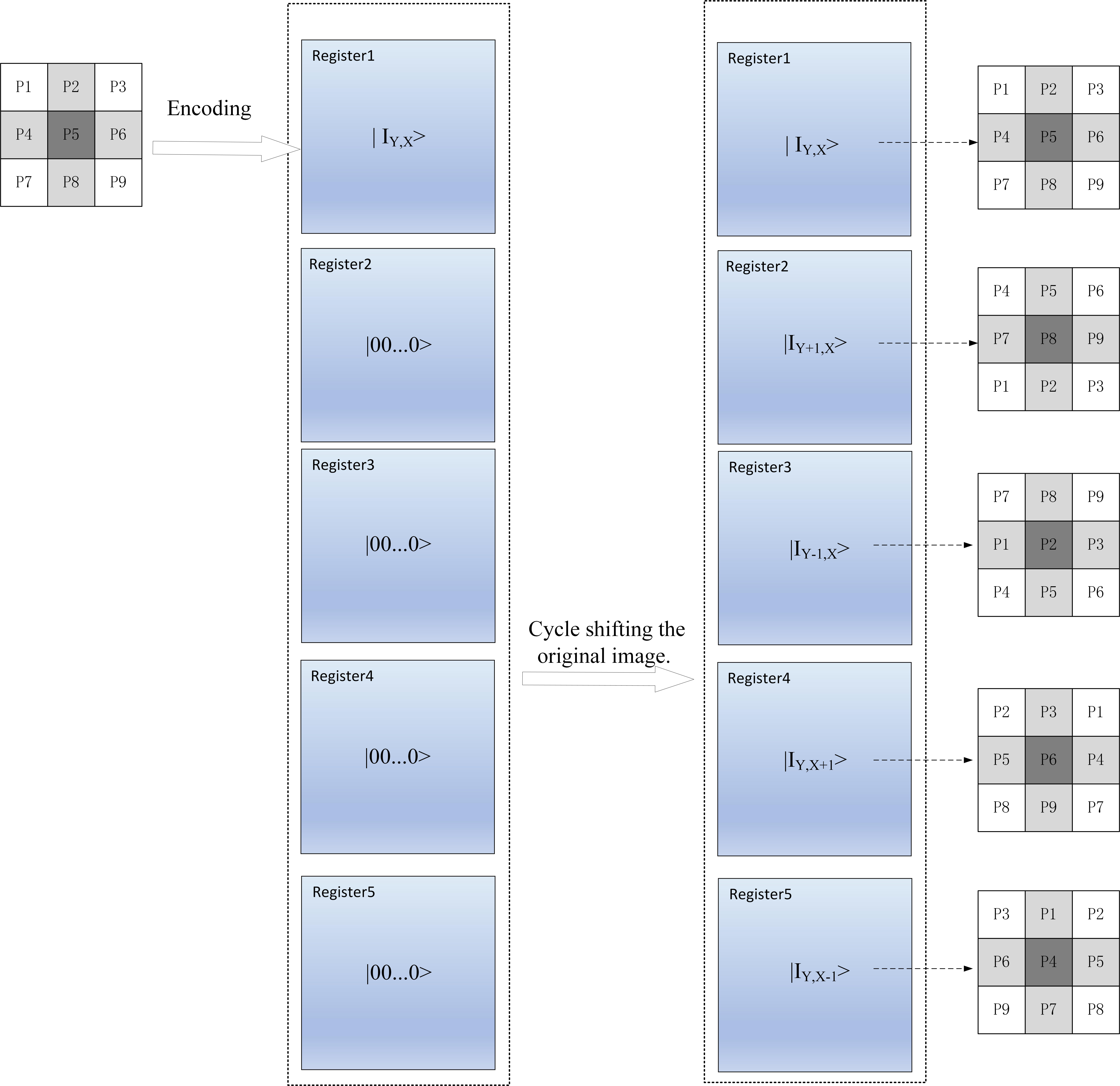}
    \caption{Schematic diagram of quantum image cyclic shift transformation.}
    \label{Fig10}
\end{figure*}

\item \textbf{Bottom hat transform/Top hat transform}

 When performing the bottom hat transformation operation, we need to dilate the image first, then erode the image, and finally subtract the original image. The top hat transformation first erode the image, then dilate the image, and finally use the original image subtract the result image of the previous step. When performing the dilation operation on the image, we need to perform a cycle shift operation on the original image according to the method in the previous step to obtain the neighborhood window pixels. Then  the QCL module is used to find the pixel with the largest grayscale in the neighborhood window. After performing the reset operation on other neighborhood pixels, we copy its maximum grayscale value to the original image (the center point of the neighborhood window) using the copy operation. This completes the dilation operation, and the specific circuit is shown in the Fig. \ref{Fig11}(a). When performing the erosion operation, we also need to perform a cycle shift operation on the original image, and then use the QCS module to find the pixel with the smallest grayscale value in the neighborhood window, and finally copy it to the original image, so that the image can be processed with erosion  operation, and the specific circuit is shown in the Fig. \ref{Fig11}(b).

\begin{figure*}
    \centering
  \subfigure[]{  \includegraphics[width=15cm]{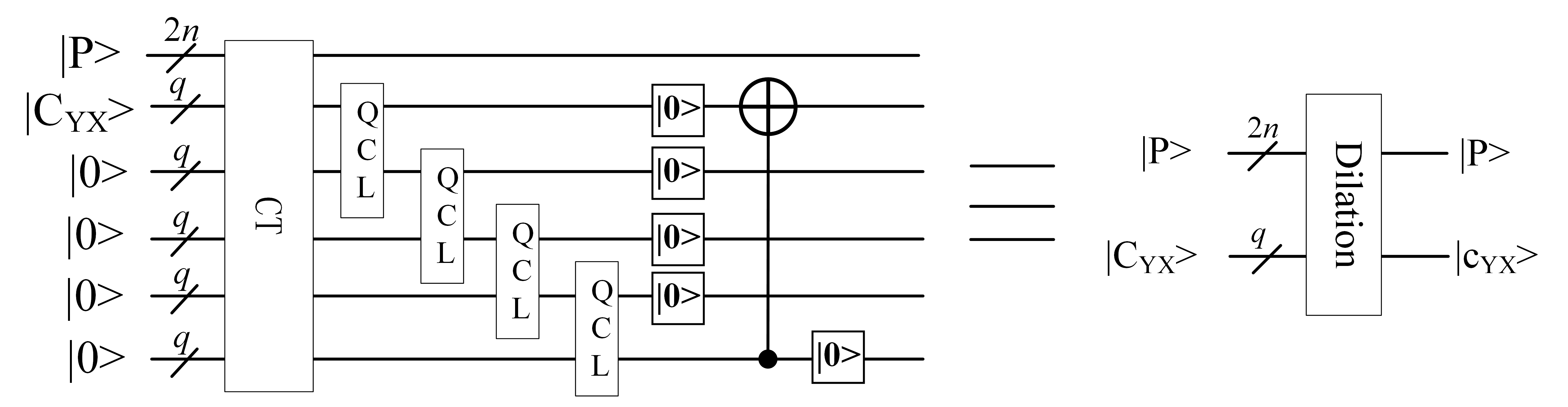}}
  \subfigure[]{  \includegraphics[width=15cm]{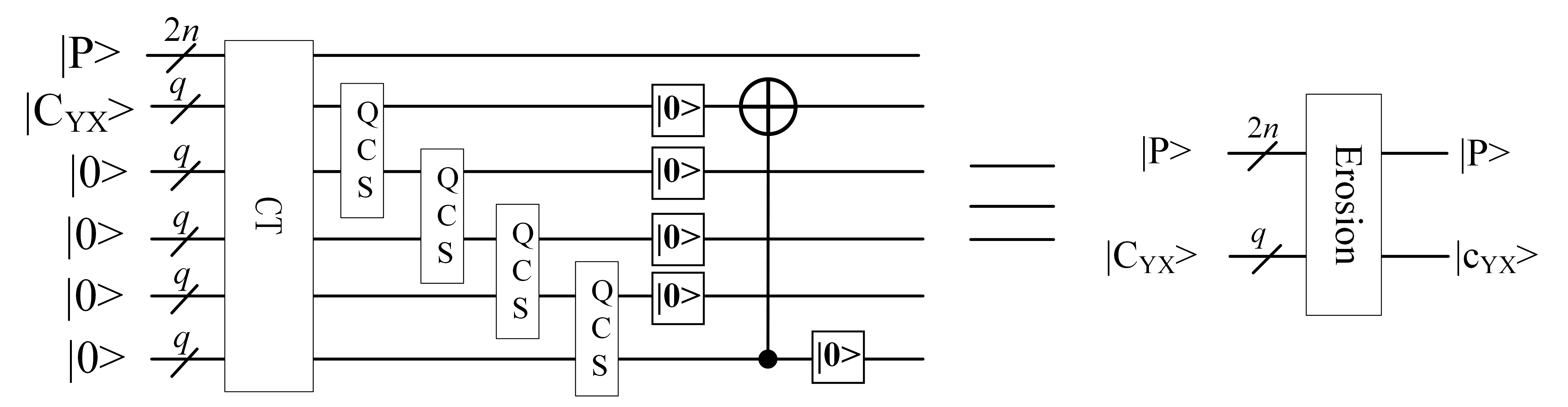}}
    \caption{The implementation circuit for dilation and erosion operations.}
    \label{Fig11}
\end{figure*}

We combine these two operations and use auxiliary qubits to replicate the original image for subsequent operations. Then, the obtained image and the original image are subtracted by the quantum subtractor, so that the bottom hat transformation or the top hat transformation can be performed. The specific quantum circuit is shown in Fig. \ref{Fig12}.

\begin{figure*}
     \centering
  \subfigure[The implementation circuit for bottom hat transformation.]{\includegraphics[width=10cm]{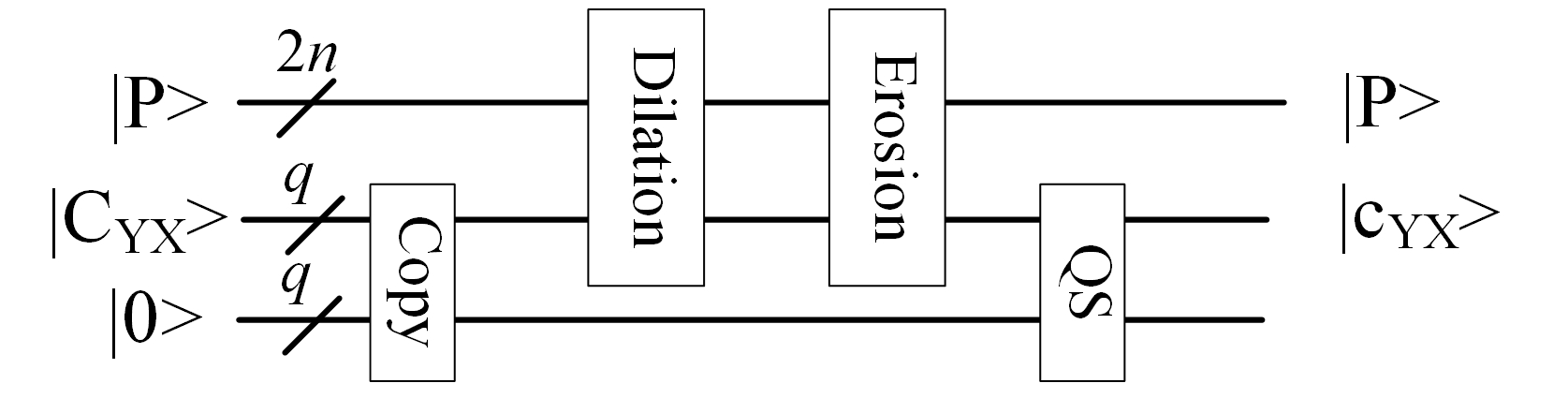}}
  \subfigure[The implementation circuit for top hat transformation.]{\includegraphics[width=10cm]{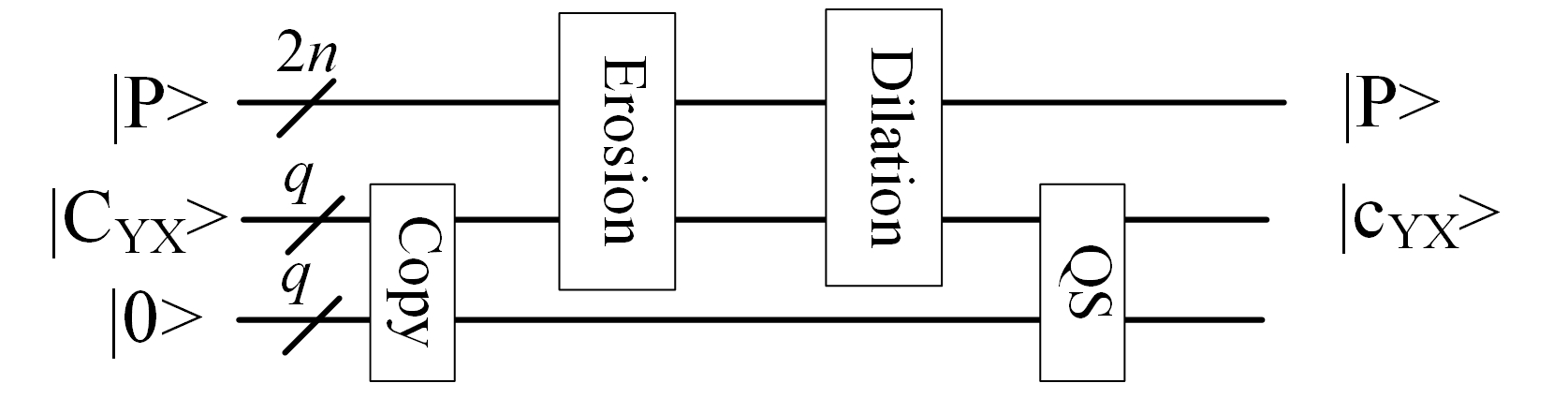}}
    \caption{The implementation circuit for bottom hat transformation and top hat transformation.}
    \label{Fig12}
\end{figure*}

\item  \textbf{  Quantum image binaryzation }

After the bottom hat transformation or the top hat transformation, we can remove the uneven lighting in the image. At this time, the binaryzation operation of the image can get better results. We use a QC to compare the grayscale value of the image after morphologically transformation with the threshold we set, and convert the grayscale value of pixels greater than or equal to the threshold to 1, and the others to 0. Since we only measure the qubit $c_0$ we need at the end, no operations are required for $c_{q-1},...,c_1$, and if the QC output $y=0$ and  $c_0=0$, then we use CNOT gate and Toffoli gate to set $c_0$ to 1. If the QC output  $y=1$ and $c_0=1$, we use CNOT gate and Toffoli gate to set $c_0$ to 0. Finally, we can get a binary image, and the specific binaryzation circuit is shown in Fig. \ref{Fig13}. The complete quantum circuits of grayscale morphological image segmentation are shown in Fig. \ref{Fig14}.

\begin{figure*}
    \centering
    \includegraphics[width=15cm]{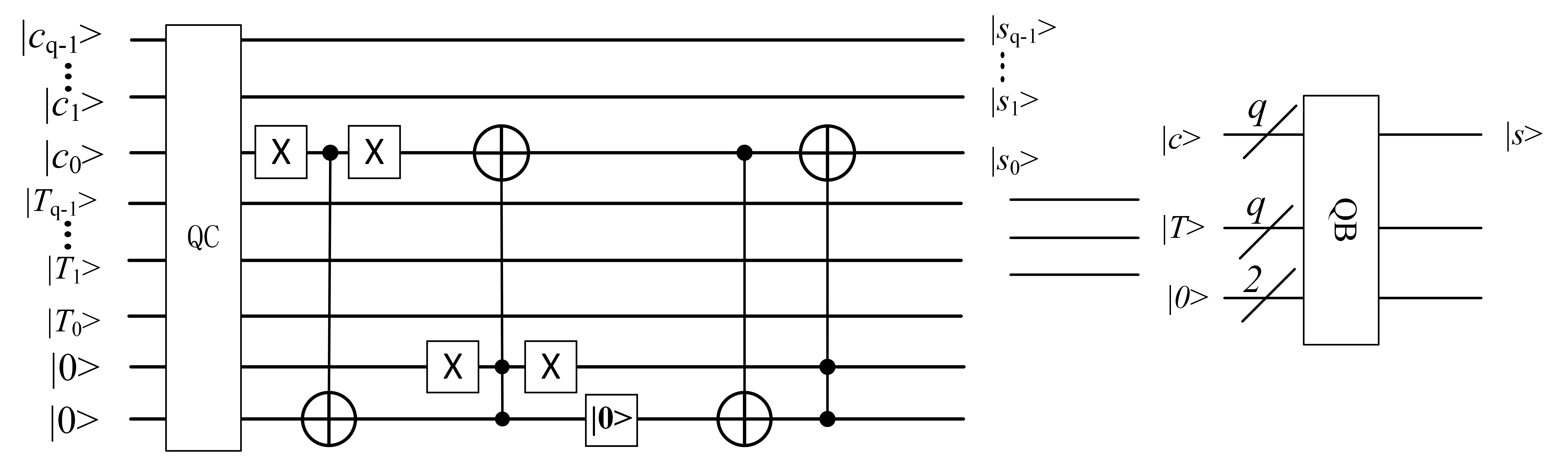}
    \caption{The implementation circuit of quantum binaryzation.}
    \label{Fig13}
\end{figure*}

\begin{figure*}
     \centering
  \subfigure[Bottom hat transformation segmentation.]{\includegraphics[width=10cm]{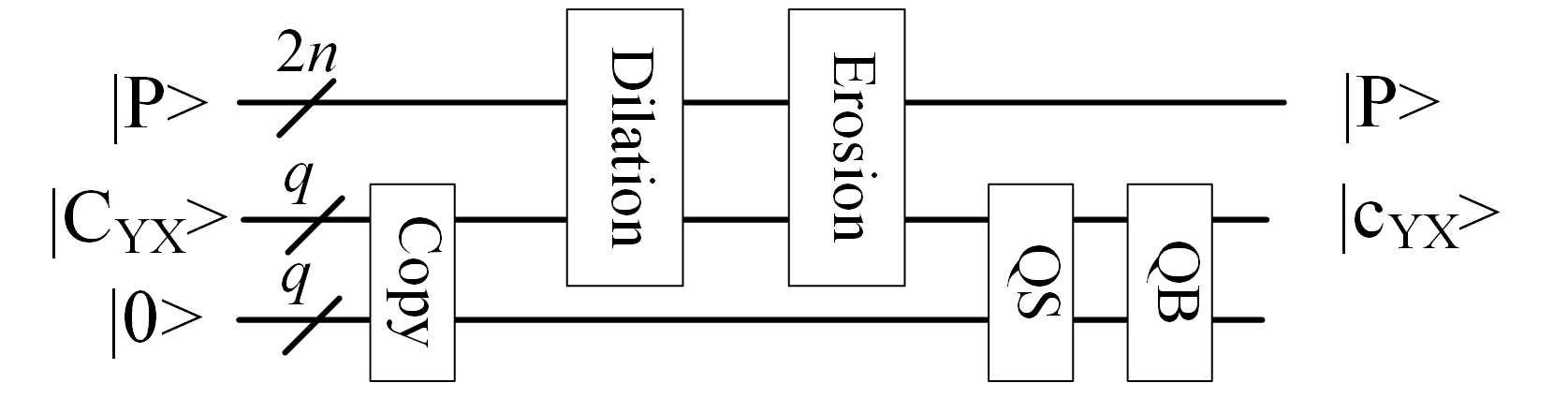}}
  \subfigure[Top hat transformation segmentation.]{\includegraphics[width=10cm]{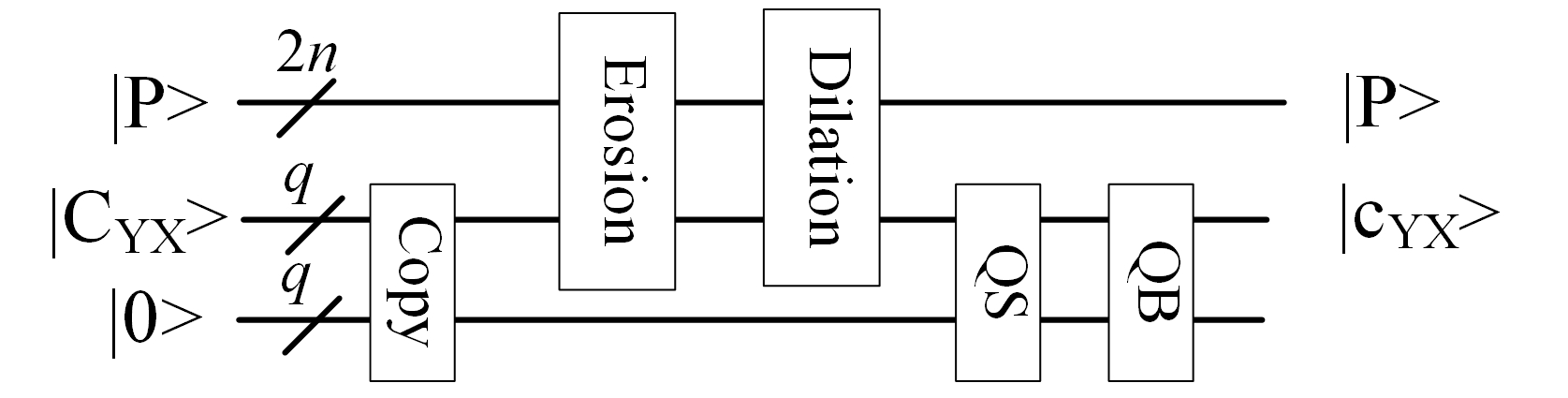}}
    \caption{The complete implementation circuits for grayscale morphological segmentation.}
    \label{Fig14}
\end{figure*}

\end{enumerate}

\section{Circuit complexity  analysis}\label{Section 4}

In quantum image processing, the complexity of the algorithm is determined by the basic quantum logic gates used in the circuit, and single-qubit gates and double-qubit gates can form arbitrarily complex quantum logic gates \cite{Nielsem2000}. Therefore, in this paper, the number of single/double-qubit quantum logic gates is used to calculate the complexity of the circuit, and their complexity are set as unit 1. Therefore, the complexity of a NOT gate, a reset gate or a CNOT gate is 1. A Toffoli gate can be composed of 5 two-qubit gates, so its complexity is 5 \cite{Nielsem2000}. The quantum cost of different quantum circuits are shown in Tab. \ref{tab3}. Taking a digital image of size $2^n \times 2^n$ and grayscale value of $[0,2^q-1]$ as an example, we will discuss the complexity of the circuit as  follows.

\begin{table}
\caption{The quantum cost and complexity of different quantum circuit.}\label{tab3}
\centering
\begin{tabular}{@{}ccc@{}}
\toprule
Module                     & Quantum Cost & Complexity \\ \midrule
Quantum comparator         & $18q-13$     & O($q$)       \\
Quantum subtractor         & $27q-43$     & O($q$)       \\
Cycle shift transformation & $n^2$           & O($n^2$)      \\
Quantum copy operation     & $q$            & O($q$)       \\
QCS/QCL                    & $21q-13$     & O($q$)       \\
Dilation/Erosion           & $n^2+7q$        & O($n^2+q$)    \\
Binaryzation               & $2q+13 $       & O($q$)       \\ \bottomrule
\end{tabular}
\end{table}

In the research of quantum image processing, we are dealing with quantum images, but since it cannot be obtained directly at this stage, it is necessary to prepare digital images into quantum images. Therefore, the complexity of the process of preparing the quantum image is not calculated in the complexity of the quantum image processing algorithm \cite{Fan2019,Chetia2021}, so we consider the complexity of this stage to be 0.

When the quantum image is cyclically shifted, we need to use the CT operation to prepare the quantum image set, and the complexity of this operation is O($n^2$) \cite{Wang2014}. In addition, we also need a Copy operation ($q$ CNOT gates) to copy the original image into auxiliary qubits for backup, which has a complexity of O($q$). So, the complexity of a CT operation is O($n^2+q$).

The quantum operations used to perform a bottom-hat or top-hat transform on an image are roughly the same, including a copy operation, a dilation operation, an erosion operation, and a subtraction operation. The dilation or erosion operation includes a cyclic shift operation, 4 QCL (QCS) operations, $q$ reset operations and 1 Copy operation (q CNOT gates). The complexity of a QCL or a QCS is O($q$), and the complexity of a reset operation and a copy operation is also O($q$). So the complexity of the dilation or erosion operation is O($n^2+q+4q+q+q$)=O($n^2+q$), and the complexity of the subtractor operation is also O($q$). So the complexity of bottom hat transformation or top hat transformation is O($q+n^2+q+n^2+q+q$)=O($n^2+q$).

The quantum binaryzation circuit includes a quantum comparator, two CNOT gates, a reset gate and two Toffoli gates. In addition, $q$ NOT gates are required to set the threshold, and the complexity of a quantum comparator is O($q$), so the overall complexity of this part is O($q+2+1+10+q$)=O($q$).

According to the above complexity analysis, we can know that the complexity of the quantum image segmentation algorithm  based on grayscale morphology is O($q+n^2+q+q$)=O($n^2+q$).  The classical grayscale morphological image segmentation needs to process each pixel separately, so its complexity is not less than O($2^{2n}$). Therefore our algorithm can achieve exponential speedup than the classical algorithm, which can  better solve the real-time problem encountered by the classical algorithm.

In addition, our algorithm is compared with other well-known segmentation algorithms. For convenience, we simplify the algorithm in \cite{Caraiman2014,Chakraborty2018} to HIS and MQCTIS, and the comparison results are shown in Tab. \ref{tab4}. HIS needs to count the grayscale value of each pixel in the image, and then select an appropriate threshold according to the histogram. Therefore, its complexity is O($\sqrt {{2^{2n}}t} $), which includes retrieval of $t$ object pixels \cite{Caraiman2014}. In order to achieve multi-level threshold of color image segmentation, MQCTIS  uses the logic of
quantum oracle or quantum black box, and the quantum oracle circuit is built using a quantum string comparator, some basic quantum
gates and a set of fredkin gates, which requires the complexity of the algorithm to be O($2^{2n}log2^{2n}$), where $2^{2n}$  is the number of pixels in the image. Therefore, the complexity of our proposed segmentation algorithm is far lower than the above segmentation algorithms.

\begin{table}
\caption{Complexity comparison of different algorithms.}\label{tab4}
\centering
\begin{tabular}{@{}ccc@{}}
\toprule
{ \textit{Algorithm}} & { \textit{Complexity}}   &  \\ \midrule
{ \textit{HIS}} \cite{Caraiman2014}       & { \textit{O($\sqrt {{2^{2n}}t} $)}}          &  \\
{ \textit{MQCTIS}} \cite{Chakraborty2018}   & { \textit{O($2^{2n}log2^{2n}$)}} &  \\
{ \textit{Proposed algorithm}} & { \textit{O($n^2+q$)}} &  \\ \bottomrule
\end{tabular}
\end{table}

\section{Experiment}\label{Section 5}
IBM opened a quantum cloud computing platform IBM Q (IBM Quantum Experience) \cite{IBM} in 2017, which includes a quantum toolkit Qiskit \cite{Qiskit}, so researchers can use Qiskit to create, compile, and run quantum computing program on quantum computers or simulators. Users can compile the Python language into the OpenQASM language through the environment created by Qiskit and anaconda, and then create a quantum circuit. After running the quantum circuit on IBM Q and measuring it, we can get the probability histogram of the qubit sequence. There are currently a total of 24 quantum computers and 5 quantum simulators on IBM Q and they support up to 8192 executions of quantum circuits. But because of the preciousness of qubits, we can only use quantum computers with 5 qubits, which is far from enough. So we can only use the quantum simulator 'ibm\_qasm\_simulator' to run our quantum circuits, which has 32 qubits and supports all the quantum logic gates we use. When used, it is mainly called by the statement “backend=IBMQ.get\_provider(‘ibm-q').get\_backend(‘ibmq\_qasm\_simulator')".

\begin{figure}
    \centering
    \includegraphics[width=5cm]{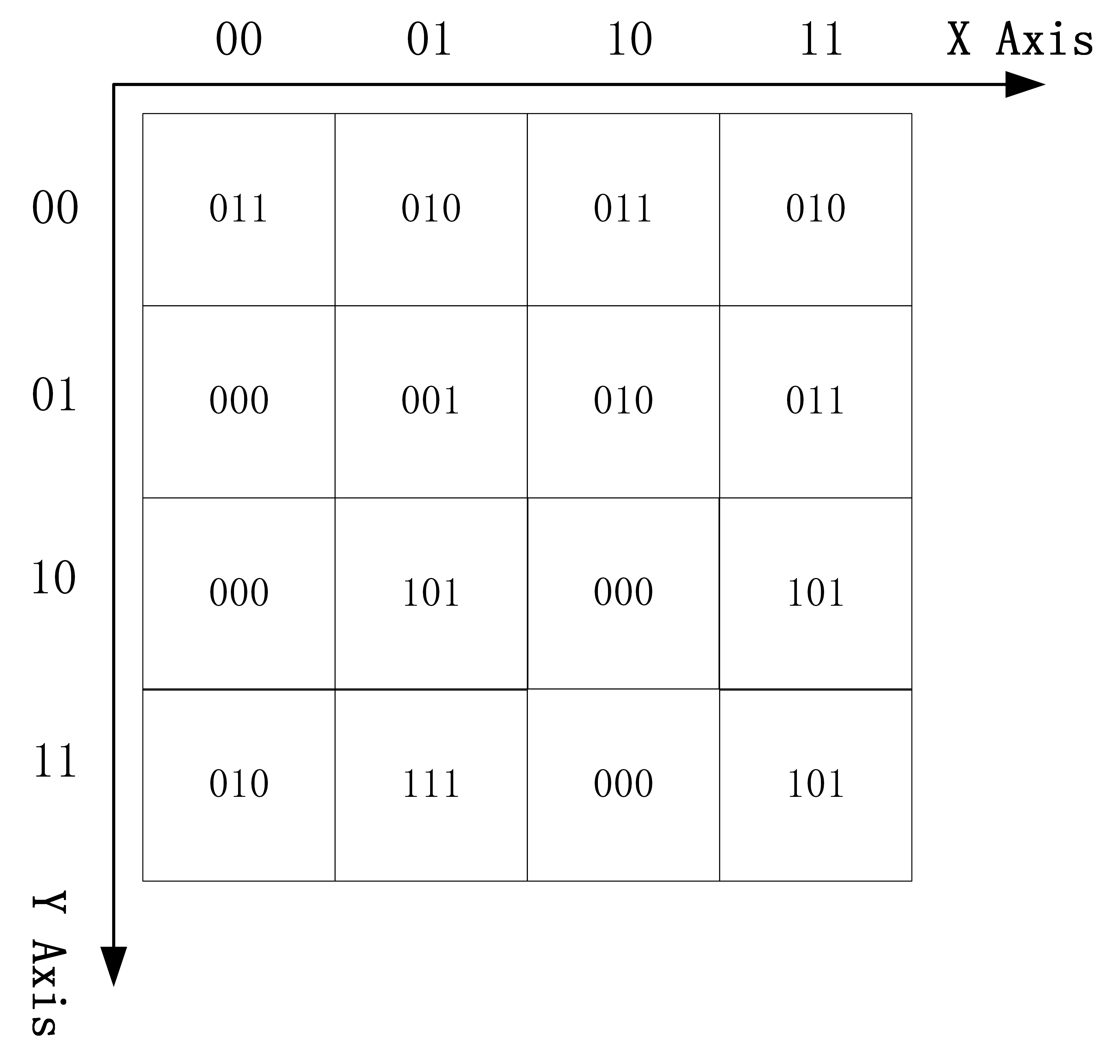}
    \caption{The Schematic of original image.}
    \label{Fig15}
\end{figure}

\begin{figure*}
     \centering
  \subfigure[The probability histogram of bottom hat transformation.]{\includegraphics[width=9cm]{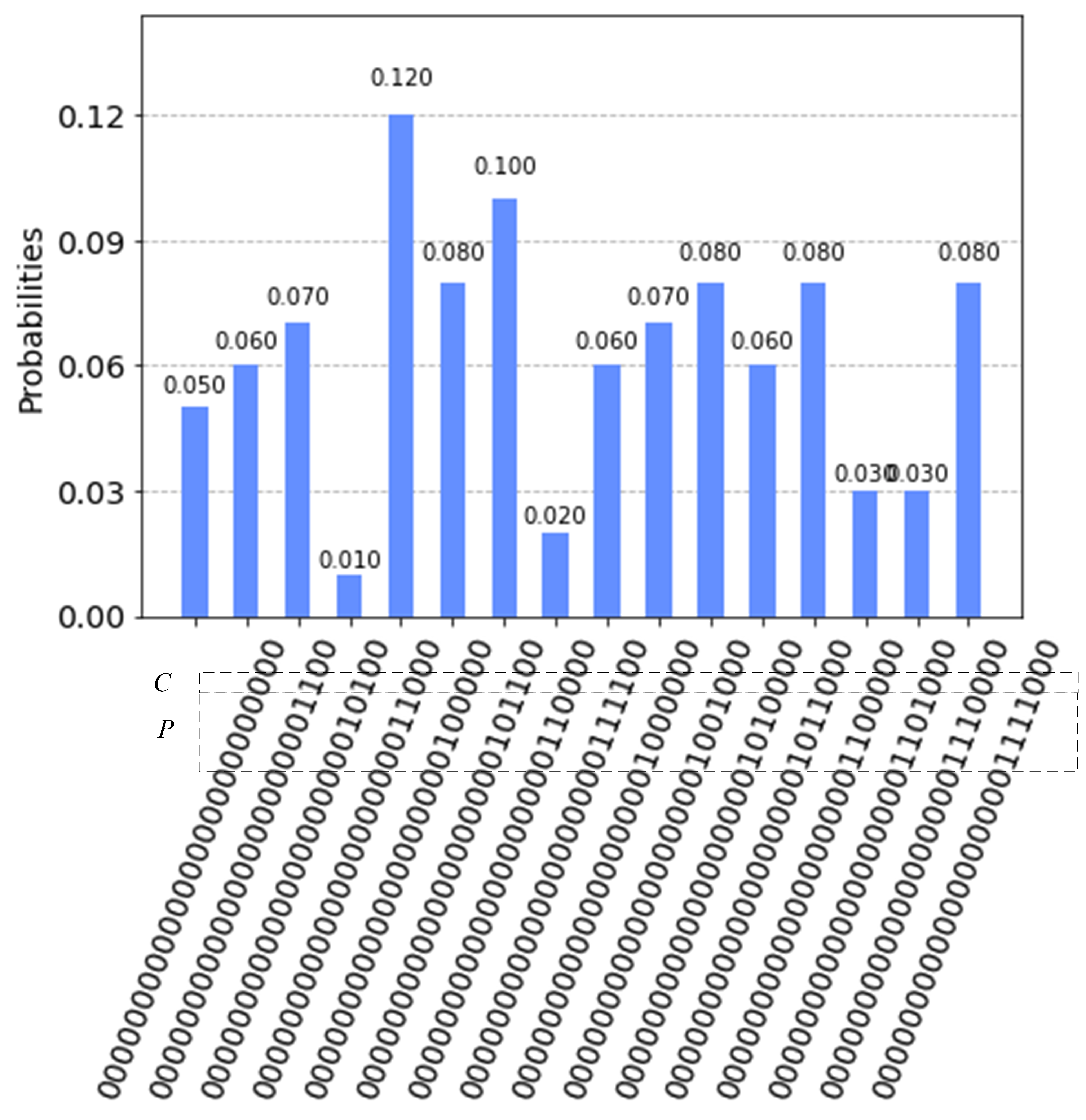}}
  \subfigure[The probability histogram of top hat transformation]{\includegraphics[width=9cm]{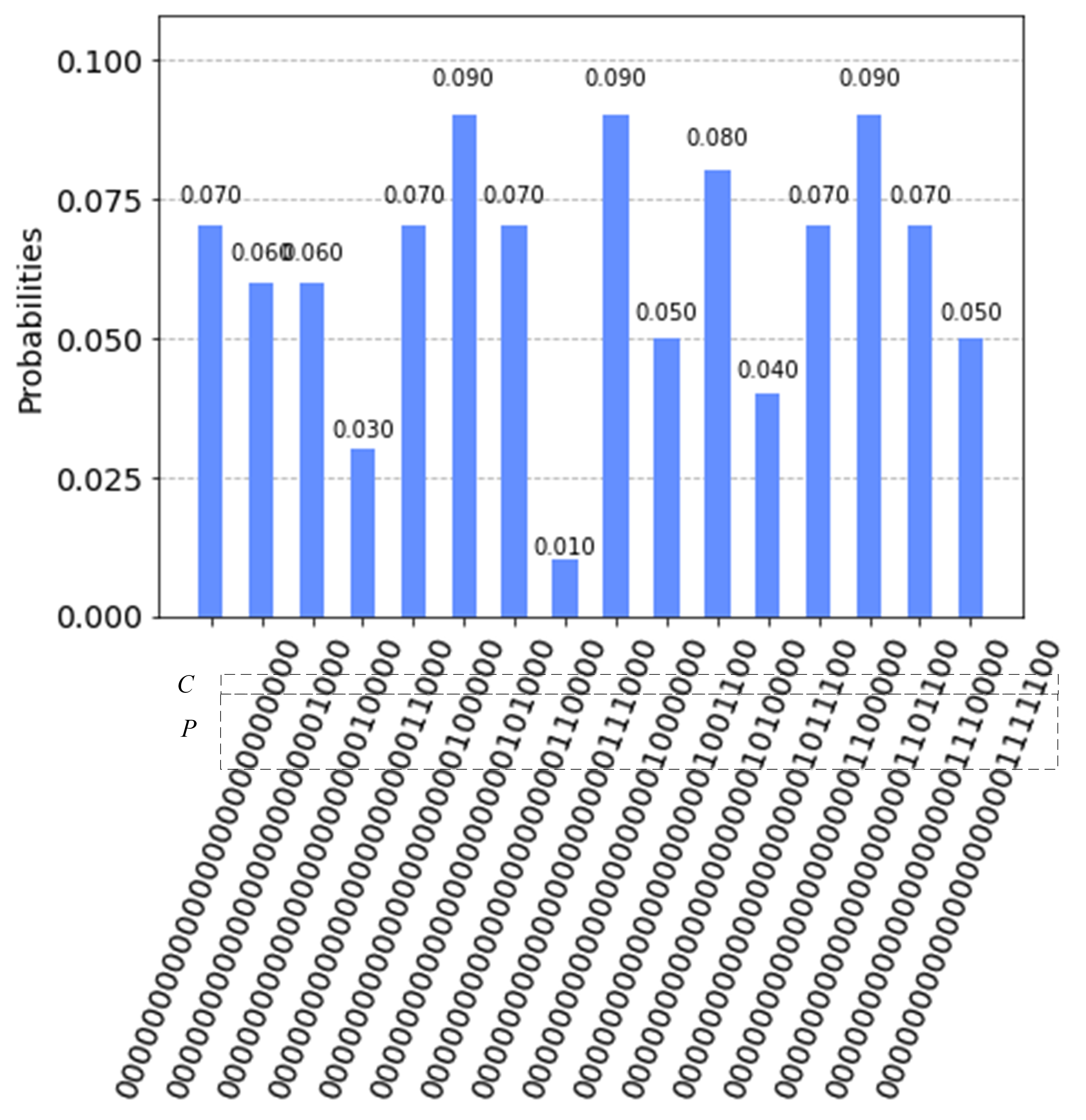}}
    \caption{The probability histograms of bottom hat transformation and top hat transformation.}
    \label{Fig16}
\end{figure*}

\begin{figure*}
     \centering
  \subfigure[The schematic diagram of the resulting image of  bottom hat transformation.]{\includegraphics[width=5cm]{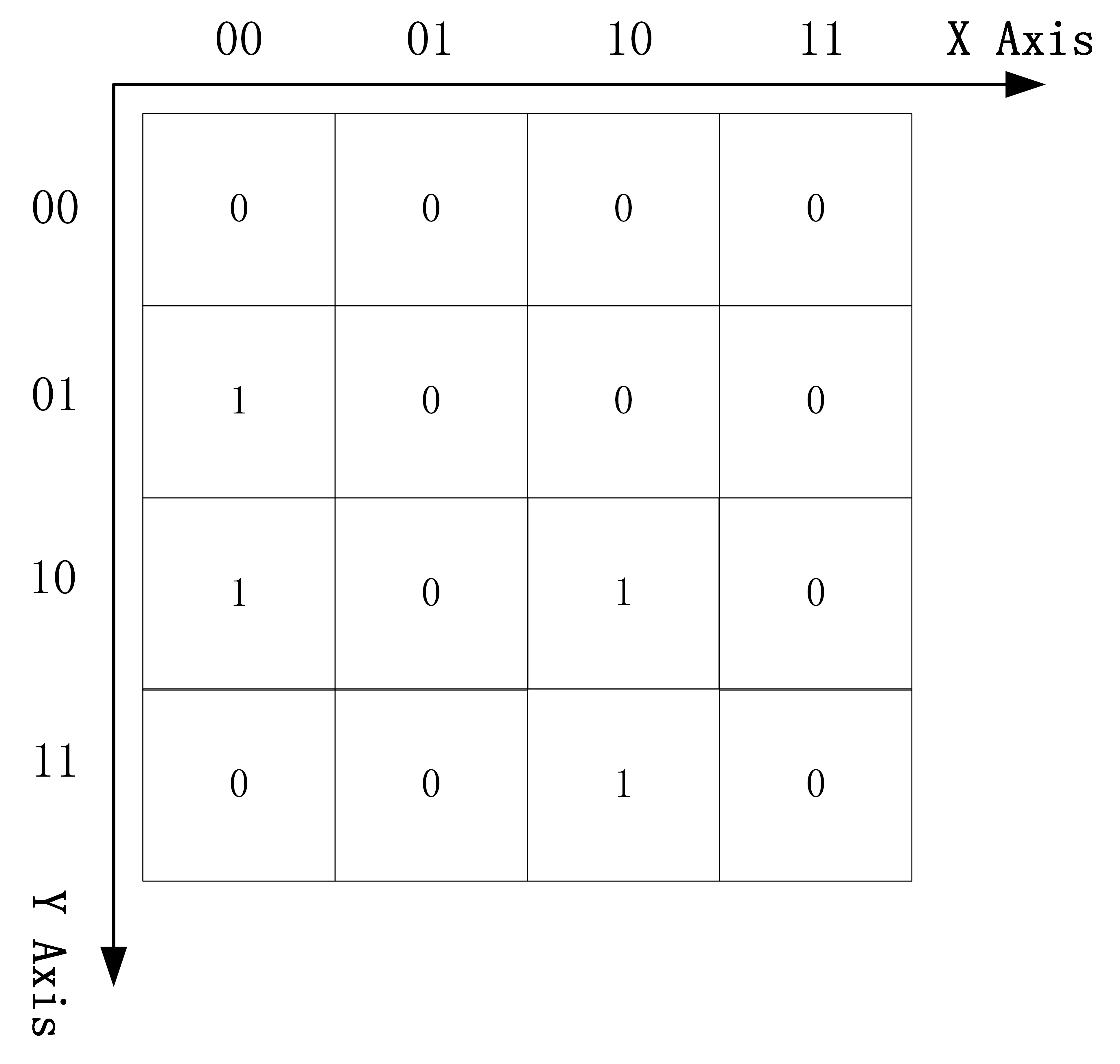}}
  \subfigure[The schematic diagram of the resulting image of  top hat transformation.]{\includegraphics[width=5cm]{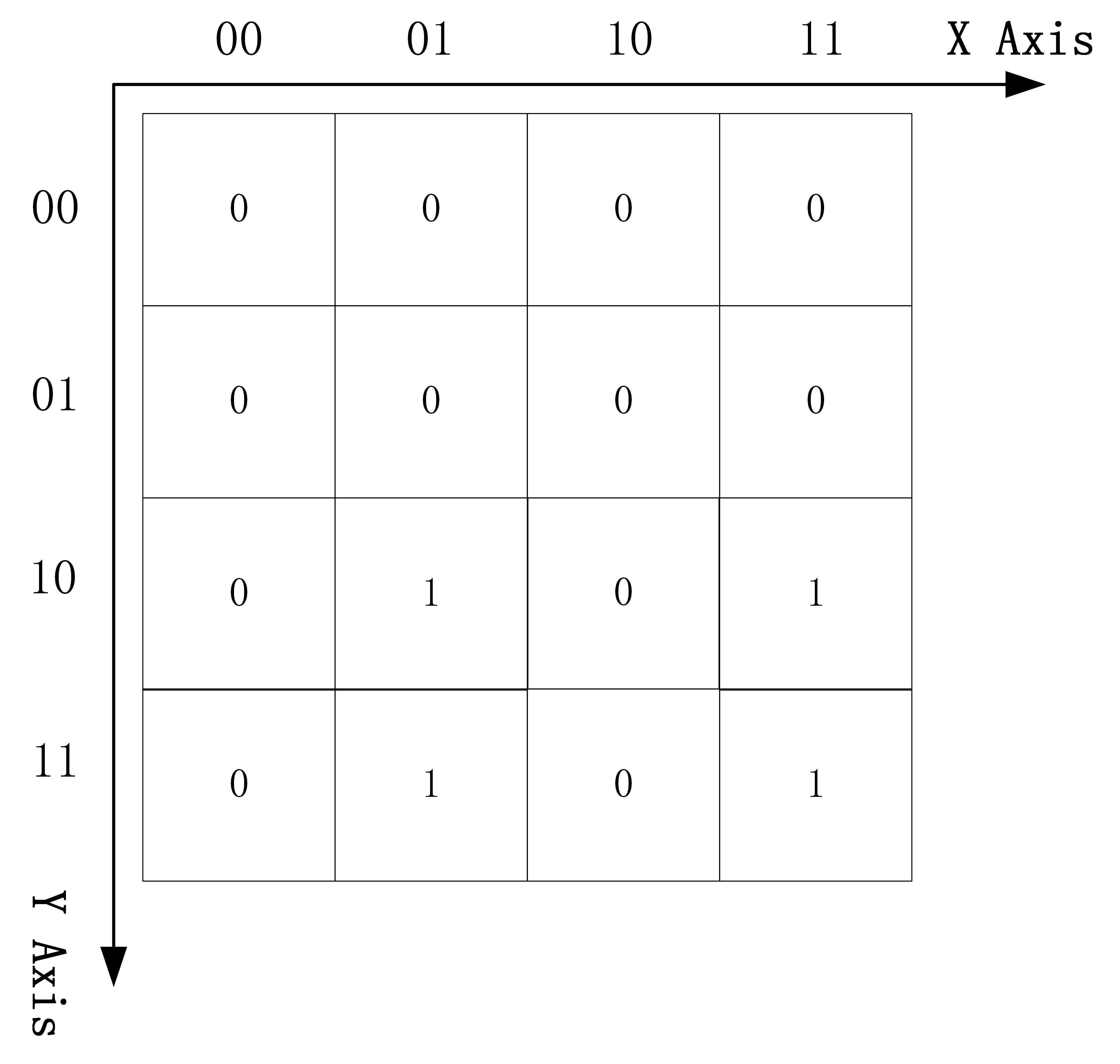}}
    \caption{The schematic diagram of the resulting image of bottom hat transformation and top hat transformation.}
    \label{Fig17}
\end{figure*}

In order to verify the feasibility of our algorithm on IBM Q, without affecting the experimental results, we selected a classical image whose size is $4\times4$ and the grayscale value is [0,7]. The schematic diagram of the image is as shown in Fig. \ref{Fig15}.  Y and X represent the coordinates of the pixel, and the grayscale value information has been marked in the figure.

According to the $4\times4$ image given above, we perform bottom hat transformation segmentation and top hat transformation segmentation on it respectively to demonstrate the feasibility of our algorithm. Since the task time of ibm\_qasm\_simulator is limited to 10000 seconds (about 2.7 hours), in order to reduce the running time of the circuit, we only measure the required qubits (the grayscale value information and the position information of the image). Figure \ref{Fig16} is the probability histogram of the resulting image, where the abscissa is the measured qubit sequence, and the ordinate is the probability of each qubit sequence. In the order from top to bottom, the third qubit is the gray value of the resulting image pixel (C=0 or 1), and the next four qubits are the corresponding position YX. The rest of the qubits are used to make up the complete circuit and we do not need to pay attention to the result. Figure \ref{Fig17} is the schematic diagram of the result image after top hat transformation and the bottom hat transformation. As can be seen from Fig. \ref{Fig16} and Fig. \ref{Fig17}, our algorithm can accurately achieve the function of grayscale morphological segmentation.

\section{Conclusion}\label{Section 6}
While various morphological algorithms are quite mature in classical image processing,  how to use quantum mechanisms to improve processing efficiency is still in its infancy. In order to solve the real-time problem of  segmenting the   uneven illumination image,  a quantum image segmentation algorithm based on grayscale morphology is firstly proposed, which uses quantum mechanism to perform grayscale morphology process on all pixels in an image at the same time,  and then quickly segment a grayscale image into a binary image. The complexity of our algorithm has been greatly improved compared with other algorithms.

In the image segmentation algorithm based on morphology,  the selection of structure elements will  influence the segmentation effect to a certain extent.  In this paper, we just consider  one simple scenario, i.e., using a fixed  cross-shaped structure element.  Therefore, how to adaptively select  the appropriate structure elements according to the image characteristics is one of our future work.


\EOD

\end{document}